\documentclass[%
 reprint,
 amsmath,amssymb,
 aps,
]{revtex4-2}

\usepackage{graphicx}
\usepackage{dcolumn}
\usepackage{bm}

\usepackage{gensymb}
\usepackage[utf8]{inputenc}
\usepackage{braket}
\usepackage{amsmath}
\usepackage{color}
\usepackage{soul}
\usepackage{xfrac}
\usepackage{amssymb}
\usepackage{comment}
\usepackage[normalem]{ulem}
\newcommand\redsout{\bgroup\markoverwith{\textcolor{red}{\rule[0.5ex]{2pt}{0.4pt}}}\ULon}

\soulregister\cite{7} 
\soulregister\ref{7} 
\soulregister\eqref{7} 

\begin{document}

\preprint{APS/123-QED}

\title{Coherent Backscattering of Entangled Photon Pairs}

\author{Mamoon Safadi$^{1}$}
\author{Ohad Lib$^{1}$}%
\author{Ho-Chun Lin$^{2}$}%
\author{Chia Wei Hsu$^{2}$}
\author{Arthur Goetschy$^{3}$}
\author{Yaron Bromberg$^{1}$}
 \email{yaron.bromberg@mail.huji.ac.il}
\affiliation{ 
$^{1}$Racah Institute of Physics, The Hebrew University of Jerusalem, Jerusalem 91904, Israel\\
$^{2}$Ming Hsieh Department of Electrical and Computer Engineering, University of Southern California, Los Angeles, California 90089, USA\\
$^{3}$ESPCI Paris, PSL University, CNRS, Institut Langevin, 1 rue Jussieu, F-75005 Paris, France
}%


\begin{abstract}
  Correlations between entangled photons are a key ingredient for testing fundamental aspects of quantum mechanics and an invaluable resource for quantum technologies. However, scattering from a dynamic medium typically scrambles and averages out such correlations. Here we show that multiply-scattered entangled photons reflected from a dynamic complex medium remain partially correlated. We observe in experiments and in full-wave simulations enhanced correlations, within an angular range determined by the transport mean free path, which prevail disorder averaging. Theoretical analysis reveals that this enhancement arises from the interference between scattering trajectories, in which the photons leave the sample and are then virtually reinjected back into it. These paths are the quantum counterpart of the paths that lead to the coherent backscattering of classical light. This work points to opportunities for entanglement transport despite dynamic multiple scattering in complex systems.
\end{abstract}

\maketitle

Entangled photons exhibit correlations that cannot be explained by classical physics. Over the past decades, physicists harnessed entangled states of photons to test some of the most peculiar predictions of quantum mechanics such as the violation of Bell's inequalities \cite{aspect1982experimental_2} and teleportation \cite{bouwmeester1997experimental}. Entangled photons have also proven to be indispensable for quantum technologies such as device independent quantum communication \cite{ekert2014ultimate} and linear optical quantum computation \cite{knill2001scheme}. While such states are indeed an invaluable resource, they are typically prone to a variety of processes that affect their nonclassical correlations. One ubiquitous and often inevitable process is light scattering from inhomogeneties. Thus, it is crucial to understand how such scattering events affect entangled photons, especially given recent rapid advances in utilizing quantum states of light in real-life scenarios such as satellite-based entanglement distribution through turbulent atmosphere \cite{yin2017satellite} and quantum imaging through biological tissue \cite{Lum:21}.
While entanglement can survive multiple scattering from a static medium if all output modes are accessible \cite{cande2014transmission,valencia2020unscrambling}, the dynamic movement of the scatterers constantly changes the state of the photons, washing out correlations in disorder-averaged states \cite{pors2011transport,ibrahim2013orbital,krenn2015twisted,leonhard2015universal}. In the classical regime, researchers have observed the existence of several so-called mesoscopic phenomena which survive the dynamic movement and disorder averaging~\cite{2006_Mishchenko_book,RevModPhys.71.313,akkermans2007mesoscopic, carminati_schotland_2021}, such as long-range correlations \cite{1988_Feng_PRL, mello88} and coherent backscattering (CBS) \cite{Kuga1984,VanAlbada1985,Wolf1985,Akkermans1986}. Thus, it is invaluable to study whether the quantum counterpart of such phenomena can exhibit correlations that are robust to disorder averaging.

Studying quantum states of multiply-scattered light poses remarkable theoretical, numerical and experimental challenges, due to the huge Hilbert space spanned by multiphoton states that occupy numerous spatial modes. To circumvent this issue, analogies of multiple scattering were studied in one-dimensional arrays of coupled single-mode waveguides, with engineered and propagation-invariant disorder. Such arrays mimic quantum walks in disordered potentials which exhibit transverse Anderson localization of photon pairs \cite{crespi2013anderson,di2013einstein,gilead2015ensemble}. However, experiments in such one-dimensional arrays do not account for the angular degrees of freedom and cannot probe mesoscopic phenomena such as CBS, universal conductance fluctuations, and universal optimal transmission. While theoretical studies indicate that the scattering of quantum states by volumetric disordered samples can exhibit diverse mesoscopic effects in the angle-resolved photon correlations \cite{beenakker2009two,ott2010quantum,cande2013quantum,cande2014transmission,schotland2016scattering}, such features could not be measured in experiments due to the low collection efficiency of multiply-scattered photons. Experiments in this regime focused instead on global features of the total reflection or transmission that do not resolve the angles \cite{lodahl2005transport,lodahl2005spatial,smolka2009observation,smolka2012continuous}. Speckles in the angle-resolved two-photon correlations, coined two-photon speckle, were measured only for thin, forward scattering diffusers, but no mesoscopic correlations were reported \cite{peeters2010observation,PhysRevA.85.033807,PhysRevA.85.033823,PhysRevLett.121.233601}.  Therefore, mesoscopic features of quantum light in high dimensions have not been observed to date.

\begin{figure*}[hbt!]
 \centering
 \includegraphics[scale=0.47]{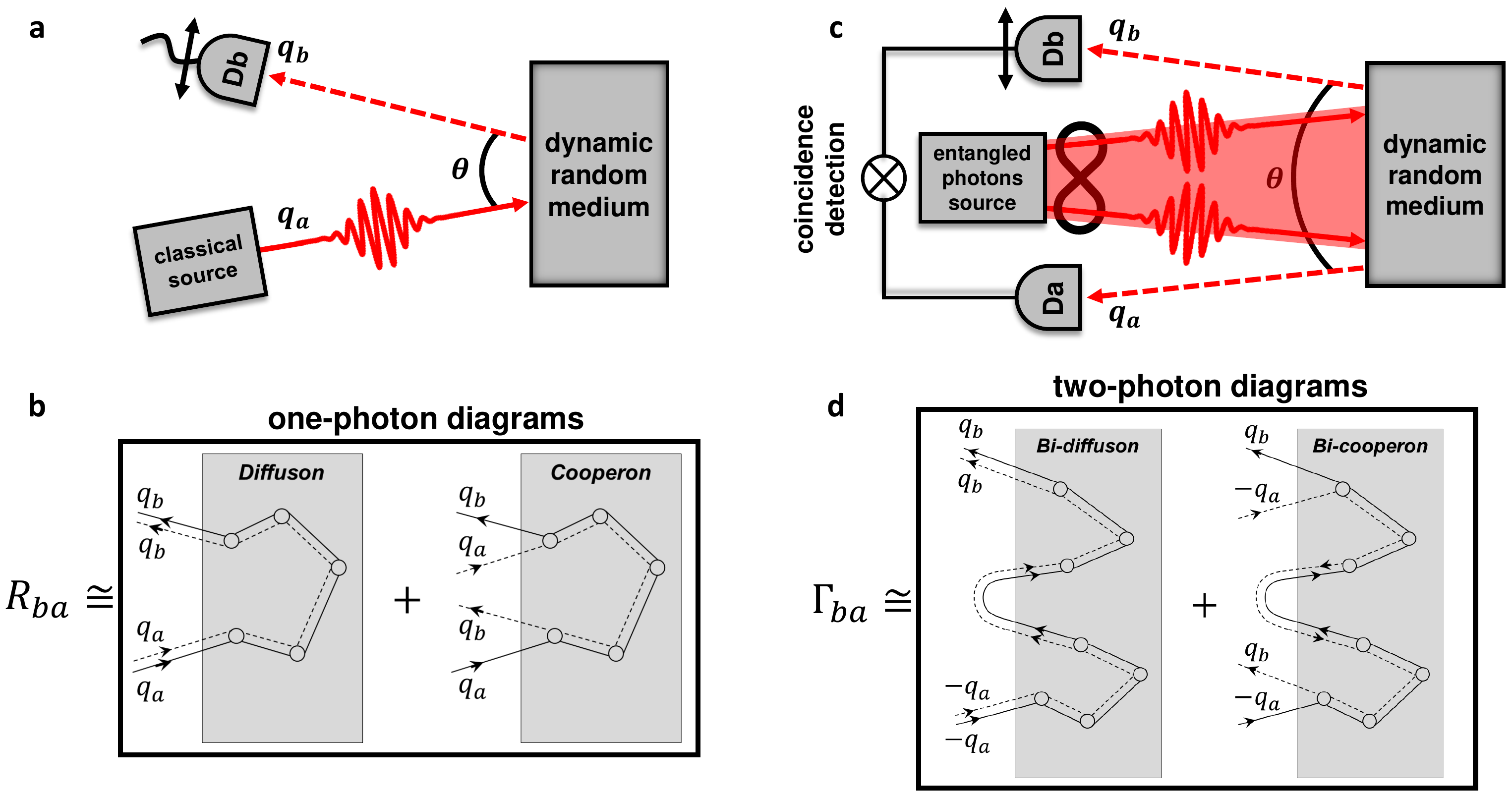}
 \caption{\textbf{Classical and quantum coherent backscattering.} \textbf{a}, Typical experimental scheme of a classical coherent backscattering experiment. A classical coherent plane-wave impinges onto a random medium, and the reflected light is collected by a detector ($D_b$). At the precise backscattering angle, $\theta=0 \degree$, an enhancement of the light is observed, coined coherent backscattering (CBS). \textbf{b}, Classical CBS diagrams of the mean reflection coefficient $R_{ba}$, where $a$ ($b$) represents the  input (output) mode with transverse momentum $\mathbf{q}_a$ ($\mathbf{q}_b$). The first term, coined diffuson, is an incoherent summation of all path pairings inside the medium. The second term, coined cooperon, contains all path pairings that visit the same scatterers but in a reversed manner. \textbf{c}, Schematic layout of a two-photon CBS (2p-CBS) experiment. A flux of entangled photon pairs illuminates the random medium and the backscattered photons with transverse momenta $\mathbf{q}_a$ and $\mathbf{q}_b$ are collected using two detectors whose coincidence events are registered. Here, $\theta$ is the angle between the two output modes. \textbf{d}, The leading diagrams found for 2p-CBS when calculating the two-photon correlation function $\Gamma_{ba}$ of the two output modes $a$ and $b$. In Klyshko's advanced wave picture, the detected mode $a$ is replaced by an illumination mode that backpropagates through the system with an opposite transverse momentum ($-\mathbf{q}_a$), and is detected in mode $b$. We coin these diagrams the bi-diffuson (first term) and bi-cooperon (second term), and they represent the two-photon generalizations of their classical counterparts (see text for more details).}
 \label{fig:1}
\end{figure*}

In this work, we experimentally and theoretically study quantum correlations between pairs of entangled photons that backscatter from a dynamic disordered sample. We discover that even after disorder averaging, the photon pairs remain correlated, revealing a new mesoscopic feature of two-photon speckle, which we coin ``two-photon coherent backscattering'' (2p-CBS). We succeeded in collecting enough photons to observe a pronounced signal by designing a scattering sample made of a thin rotating diffuser followed by a mirror that reflects the photons back through it. This double-passage configuration allows the photons to scatter twice from the same position on the diffuser. We also perform a theoretical and numerical study in a true multiple-scattering medium to identify the fundamental scattering processes at the origin of the 2p-CBS. Our analysis reveals that the experimentally observed correlations are universal and not unique to the double-passage configuration considered in experiment. Finally, we show that one can achieve a better estimation of the transport mean free path using 2p-CBS compared to classical CBS.

When a classical plane-wave illuminates a disordered medium, a two-fold enhancement of the backscattered intensity is observed in the direction opposite to the incoming wave ($\theta = 0 \degree$ in Fig.~1a). This region of enhanced intensity, coined the CBS cone, is revealed only after ensemble averaging over realizations of the disorder, as a single realization is dominated by a strongly fluctuating speckle. Classical CBS has been studied by scattering laser light off a wide array of disordered samples, including powder suspensions \cite{Kuga1984,Wolf1985,VanAlbada1985,Akkermans1986}, random phase screens \cite{Jakeman1988,dogariu1995enhanced,schwartz2005enhanced}, amplifying random media \cite{PhysRevLett.75.1739}, biological tissues \cite{Yoo1990}, human bones \cite{Derode2005}, multimode fibers \cite{Bromberg2016}, ultracold atom gases \cite{Labeyrie1999}, and liquid crystals \cite{Huang2001}. CBS was also observed for other types of classical light, such as pseudothermal \cite{Scalia_2013} and Raman light \cite{Fazio2017}, as well as for other types of waves such as acoustic \cite{bayer1993weak}, seismic \cite{larose2004weak}, and quantum matter waves \cite{jendrzejewski2012coherent}. The universality and robustness of CBS arises from the optical reciprocity principle \cite{potton2004reciprocity}, as revealed by a diagrammatic decomposition of the field over the scattering paths, featuring  two dominant terms of the reflected intensity \cite{Akkermans1986}. The first term, the \textit{diffuson}, accounts for a homogeneous background  and corresponds to all pairs of paths propagating through the same scatterers identically. The second term, the \textit{cooperon}, accounts for the enhancement observed in the backscattering direction and corresponds to constructive interference of all path pairs propagating through the same scatterers but in a reversed order (Fig.~1b).

\begin{figure*}[hbt!]
 \centering
 \includegraphics[scale=0.47]{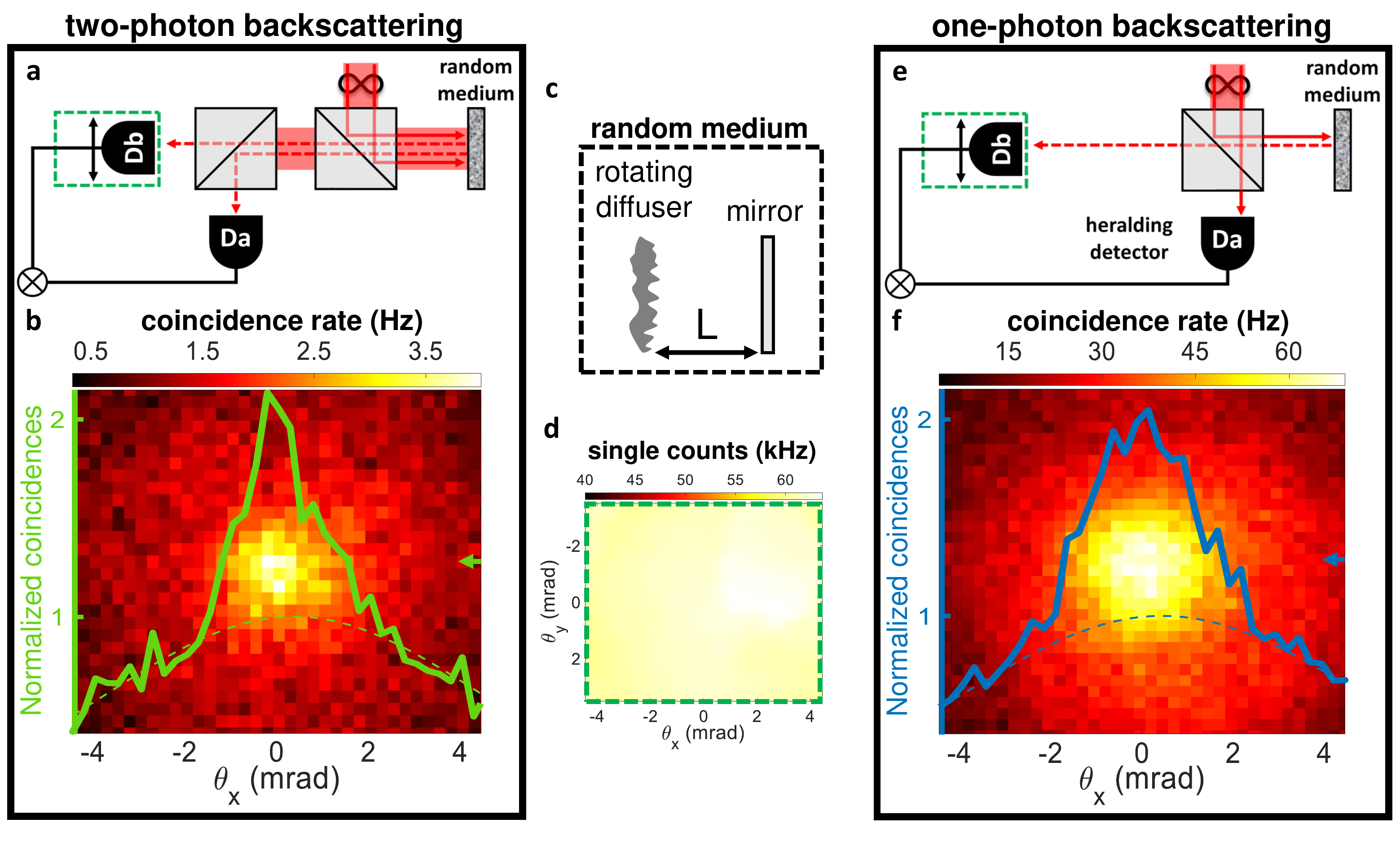}
 \caption{\textbf{Experimental observation of two-photon and one-photon coherent backscattering}. \textbf{a}, Scheme of the 2p-CBS experiment where entangled photon pairs illuminate the random sample. The backscattered photons are measured via coincidence logic of detectors $D_a$ and $D_b$. \textbf{b}, The 2D coincidence map observed and a normalized cross section along the indicated row (green arrow). The coincidence distribution exhibits a clear enhancement in the backscattering direction. \textbf{c}, The random sample consists of a rotating diffuser and a plane mirror placed behind it at a distance $L$. \textbf{d}, The single counts registered by the scanning detector $D_b$ exhibit a homogeneous distribution over the scanned region. \textbf{e}, Scheme of the 1p-CBS experiment, which mimics classical CBS. Heralded single photon illumination is obtained by a coincidence logic between detector $D_b$ and detector $D_a$ which is now placed before the random sample. \textbf{f}, The 2D coincidence map observed and a normalized cross section along the indicated row (blue arrow). Once again, the coincidence distribution exhibits a clear enhancement in the backscattering direction. Despite the diffuser-mirror spacing being the same ($L=2.5$ cm), the 2p-CBS shape is narrower than the 1p-CBS shape. Both experiments are performed at the far-field plane of the random medium. The transverse momenta of the coincidence counts as well as the single counts distribution are expressed in terms of angles $\boldsymbol{\theta}_{x,y}=\boldsymbol{q}_{x,y}/k$, where $k$ is the photon wavenumber. The dashed lines in panels \textbf{b} and \textbf{f} indicate the background fit for each of the 2p-CBS and 1p-CBS experiments to which we perform the normalization. More details about the setup and the fitting procedure are given in Methods and the Supplementary Sections 1 and 4.}
 \label{fig:2}
\end{figure*}

The question of whether entangled photon pairs also exhibit coherent backscattering has not been addressed to date. Quantum interference of photon pairs is probed by measuring the rate of coincidence events, namely simultaneous detection of two photons by two single-photon detectors (Fig.~1c). A convenient interpretation of such detection scheme is provided via Klyshko's advanced wave picture \cite{Klyshko}, in which the joint two-photon probability of detecting a photon in mode $a$ and a photon in mode $b$ is mapped to the probability of a photon launched into the system from mode $a$ and detected in mode $b$, after traversing the optical system twice \cite{WALBORN201087}. Using this representation and a rigorous diagrammatic expansion, we will show (see discussion below and Supplementary Sections 3 and 5) that two-photon interference of the backscattered light does not vanish after disorder averaging and is governed by two new leading diagrammatic terms. The first term, which we coin the \textit{bi-diffuson}, corresponds to all path pairs that visit the same scatterers inside the medium, leave it, and are virtually reinjected into it once more undergoing another scattering sequence. The second term, which we coin the \textit{bi-cooperon}, contains all the path pairs that visit the same scatterers in the first and second passages but in a reversed order (see Fig. 1d). 

To experimentally study the backscattering of entangled photons, we illuminate the rotating random sample with a stream of spatially entangled photon pairs generated by spontaneous parametric down conversion (SPDC) (Fig.~2a and Supplementary Fig.~S1). The photon pairs are generated at the same wavelength of $\lambda = 808$ nm and the same polarization (see Methods for more details). In the thin crystal regime \cite{WALBORN201087}, the SPDC photons are maximally entangled in their spatial degree of freedom and can be described by an Einstein-Podolsky-Rosen (EPR) state $\ket{\psi} =\frac{1}{\sqrt{2N}} \sum_{i=1}^{N} \hat{c}_{\mathbf{q}_i}^{\dagger}\hat{c}_{-\mathbf{q}_i}^{\dagger}\ket{0}$, where $N$ is the number of modes illuminating the random sample and $\hat{c}_{\mathbf{q}_i}^{\dagger}$ is the creation operator of an incident mode with transverse momentum  $\mathbf{q}_i$. We then measure the disorder-averaged coincidence rate between detectors $D_a$ and $D_b$, placed at the far-field of the rotating sample (Fig.~2a). The temporal coincidence window used to register simultaneous arrival of photons to the detectors is chosen to be $800$ ps, much shorter than the average separation time between detected pairs (a few $\mu$s). To overcome the challenge of low collection efficiencies of photons backscattered from multiple-scattering samples, we implement the double-passage configuration using a thin diffuser with a scattering angle of $\theta_0 \approx 4.4$ mrad, and a mirror at a distance $L$ behind it (Fig.~2c). This choice of random medium allows us to achieve a collection efficiency of near unity using low numerical aperture lenses (NA $\approx$ 0.1), relaxing the need to use high NA objectives that introduce multiple back-reflections which typically overwhelm the CBS effect. 

The coincidence map of detectors $D_a$ and $D_b$ reveals a sharp 2-to-1 enhancement in the backscattering direction (Fig.~2b). To our knowledge, this is the first observation of coherent backscattering of a quantum state of light, which we refer to as 2p-CBS. In contrast, the spatial distribution of the photons detected by the single counts of $D_b$ reveals no discernible structure (Fig.~2d). The fact that the backscattering enhancement is observed only in the two-photon correlation map and not in the single counts distribution is a hallmark of two-photon interference, distinguishing it from classical CBS. We note that in order to obtain a prominent enhancement of the non-classical light, one has to surpass the two main noise sources in experiment: the speckle noise of the disorder and the Poisson noise of the coincidence events. To surpass the Poisson noise, we integrated each pixel in the coincidence map for over 200 seconds to accumulate hundreds of coincidence events per pixel. To surpass the speckle noise, the diffuser was rotated and translated in the transverse direction, covering its entire surface, during the acquisition time. This way, we were able to achieve a sufficient signal-to-noise ratio and observe a pronounced peak of the backscattered quantum light.

We now compare the features of the 2p-CBS shape to the classical CBS shape. Since the flux of classical light is identical to that obtained by repeatedly illuminating the sample with a stream of single photons, classical CBS is identical to the CBS obtained using a single photon source \cite{Smolka2010}. Here, we use heralded single photons to measure the classical CBS, referred to as 1p-CBS, by placing the static detector $D_a$ before the scattering sample (Fig.~2e). Now, detection of a photon by detector $D_a$ heralds the presence of its twin and collapses its state to a photon in a plane-wave mode with a well-defined transverse momentum which illuminates the rotating sample. This backscattered photon is then collected with detector $D_b$, and its coincidence counts with detector $D_a$ are recorded. Figure~2f depicts the observed 1p-CBS coincidence map, which is clearly wider than the 2p-CBS.

To investigate the structure of the enhanced region in 2p-CBS, we calculate the two-photon correlation function, $\Gamma_{ba}$, defined as
\begin{equation} \label{Gamma}
\Gamma_{ba} = \overline{\braket{\psi|:\hat{n}_{\mathbf{q}_a}\hat{n}_{\mathbf{q}_b}:|\psi}}~,
\end{equation}
where $\hat{n}_{\mathbf{q}}=\hat{d}_{\mathbf{q}}^{\dagger}\hat{d}_{\mathbf{q}}$ is the photon number operator of a reflected mode with transverse momentum $\mathbf{q}$ and $\hat{d}_{\mathbf{q}}^{\dagger}$ ($\hat{d}_{\mathbf{q}}$) being the corresponding creation (annihilation) operator, $:(\dots):$ stands for normal ordering, and $\overline{\mbox{\dots\rule{0pt}{1.5mm}}}$ represents ensemble averaging over different realizations of disorder. This correlation function is proportional to the coincidence rate of two detectors measuring photons reflected by the sample with transverse momenta $\mathbf{q}_a$ and $\mathbf{q}_b$ (see Supplementary Section 2).
The operators of the incoming and outgoing modes are related through the reflection matrix $r$ of the sample, as $\hat{d}_{\mathbf{q}_a}= \sum_{a'} r_{\mathbf{q}_a,\mathbf{q}_{a'}}\hat{c}_{\mathbf{q}_{a'}}$. Inserting the reflection matrix and the EPR state into Eq.~\eqref{Gamma}, we obtain
\begin{equation} 
    \Gamma_{ba} = \frac{2}{N} \overline{|\sum_i r_{\mathbf{q}_b,\mathbf{q}_{i}}r_{\mathbf{q}_a,-\mathbf{q}_{i}}| ^2} = \frac{2}{N} \overline{\left|(r^2)_{\mathbf{q}_b,-\mathbf{q}_{a}}\right|^2}~,  \label{eq_Gamma_r2}
\end{equation}
where in the second equality we used optical reciprocity $r_{\mathbf{q},\mathbf{q}'}=r_{-\mathbf{q}',-\mathbf{q}}$ to express $\Gamma_{ba}$ by the matrix product $r^2$. 

The 2p-CBS can now be computed by decomposing $r^2$ over the scattering paths. Modeling scattering by the thin diffuser as Gaussian random processes, we find that in the limit of large diffuser-mirror spacing, $kL \theta^2_{0}\gg1$ ($k=2\pi/\lambda$ being the photon wavenumber), $\Gamma_{ba}$ is dominated by (Supplementary Section 3)
\begin{multline} \label{eq:3}
    \Gamma_{ba} \propto \left\{ 1+\exp\left[-\frac{\left(2L\theta_{0}\right)^{2}}{2}\left(\mathbf{q}_{a}-\mathbf{q}_{b}\right)^{2}\right]\right\}\\\times\exp\left[-\frac{\left(\mathbf{q}_{a}+\mathbf{q}_{b}\right)^{2}}{4k^{2}\theta_{0}^{2}}\right]. 
\end{multline}
The first term in the curly brackets can be interpreted as the bi-diffuson (the homogeneous background) while the second can be interpreted as the bi-cooperon (the enhanced region). Since the scattering is anisotropic, an envelope representing the finite scattering angle of the diffuser multiplies both terms.

The 1p-CBS, meanwhile, is given by (Supplementary Section 3):
\begin{multline} \label{eq:4}
    R_{ba} \propto \left\{ 1+\exp\left[-\frac{\left(L\theta_{0}\right)^{2}}{2}\left(\mathbf{q}_{a}+\mathbf{q}_{b}\right)^{2}\right]\right\}\\\times\exp\left[-\frac{\left(\mathbf{q}_{a}-\mathbf{q}_{b}\right)^{2}}{2k^{2}\theta_{0}^{2}}\right]. 
\end{multline}
$R_{ba}$ is the mean reflection coefficient that represents the intensity scattered from mode $\mathbf{q}_a$ to mode $\mathbf{q}_b$, where the first term in the brackets represents the background that corresponds to the diffuson, and the second term represents the CBS shape and corresponds to the cooperon. Once again, an envelope term accounting for the finite scattering angle of the diffuser multiplies both terms. 

\begin{figure}[hbt!]
 \centering
 \includegraphics[scale=0.23]{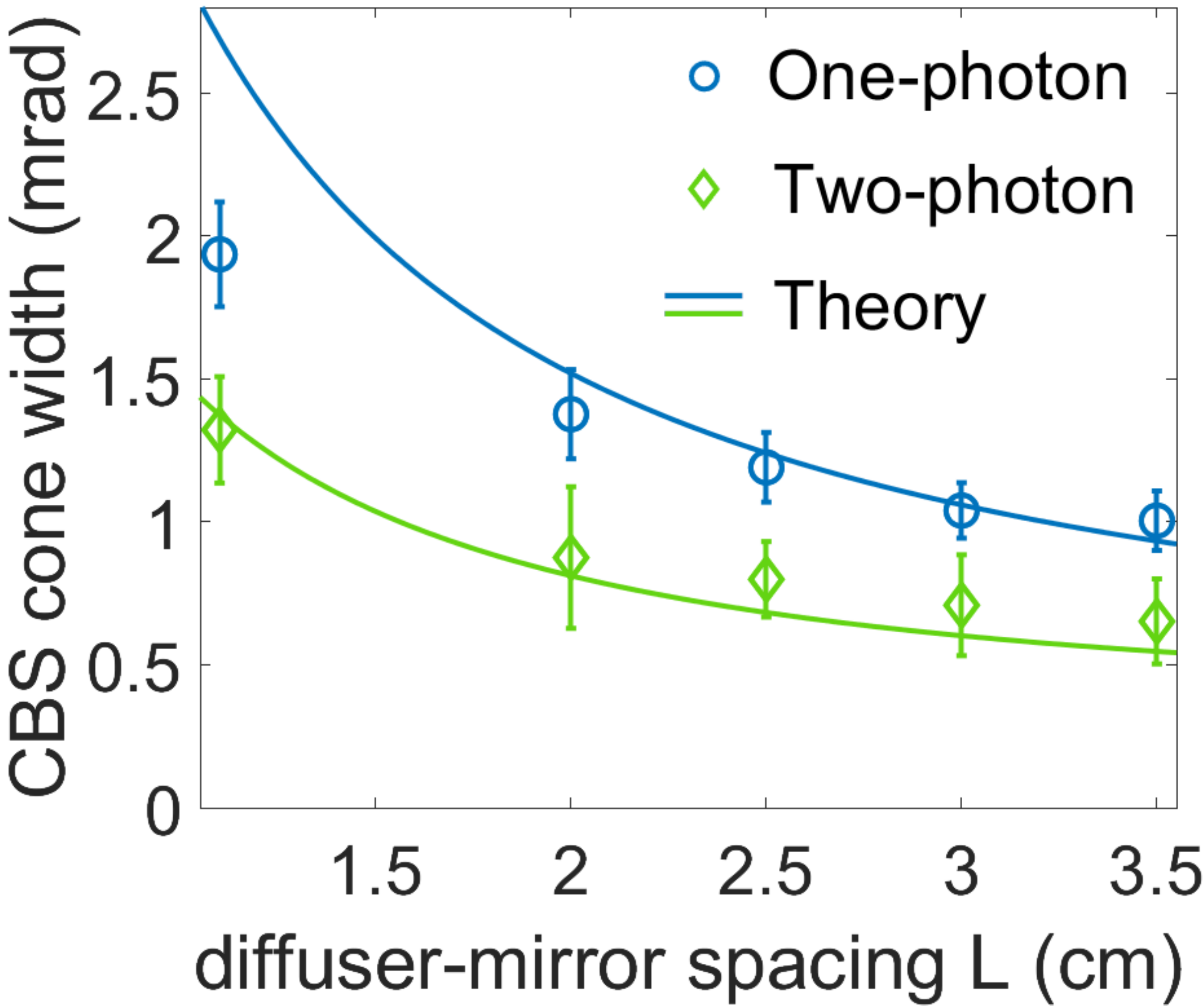}
 \caption{\textbf{2p-CBS and 1p-CBS widths against the diffuser-mirror spacing L}. The experimentally measured CBS widths of the 2p-CBS are in green diamonds and of the 1p-CBS in blue circles. Solid green line corresponds to the predicted width for 2p-CBS $\Delta \theta_{2p} = \sqrt{\delta_{2p}^2 + (2kL\theta_0)^{-2}}$, where $\delta_{2p}^2=2\sigma^2/f_4^2$ accounts for the angular resolution in the 2p-CBS experiment, determined by the radius of the fiber-coupled detectors ($\sigma=50$ $\mu$m) and the focal length of the far-field lens ($f_4=200$ mm). Solid blue line corresponds to the theoretically predicted 1p-CBS width, $\Delta \theta_{1p} = \sqrt{\delta_{1p}^2 + (kL\theta_0)^{-2}}$, where $\delta_{1p}^2=\sigma^2/f_1^2+\sigma^2/f_4^2$ is the angular resolution in the 1p-CBS experiment, determined by the divergence of the heralding single photon, located at the far-field of the crystal ($f_1 = 150$ mm), and the angular resolution of the fiber-coupled detector. Error bars indicate the confidence intervals of the fit parameter for the CBS width. See Supplementary Section 3 for more information. 
 }
 \label{fig:3}
\end{figure}

We then measured the 2p-CBS and 1p-CBS angular widths as a function of the diffuser-mirror spacing $L$. The experimental widths depicted in Fig.~\ref{fig:3}  (green diamonds for 2p-CBS and blue circles for 1p-CBS) agree with the theoretical curves (solid curves). At small distances, deviations become more apparent, and we attribute this to the CBS profile being washed out by the finite background. In the Supplementary Section 4, we provide a detailed description of the measuring process and fitting procedure. We note that as the angular widths of the 2p-CBS and 1p-CBS shapes scale like $(2kL\theta_0)^{-1}$ and $(kL\theta_0)^{-1}$ respectively, it signifies that the shape of the 2p-CBS corresponds to the shape of the 1p-CBS yet with a two times larger wavenumber. Interestingly, this unique feature of entangled photons mimicking single-photons at double the wavenumber---a quantum feature of entangled photons that often leads to super-resolution and super-sensitivity \cite{boto2000quantum}---appears to survive scattering and disorder averaging. 

We now address the important question of the universality of the 2p-CBS. Is this phenomenon restricted to the double-passage configuration and washed out in the presence of strong multiple scattering, or is it robust against any type and strength of disorder? To answer this question, we consider the propagation of entangled photons in diffusive opaque media. In such systems, light experiences on average a random walk, and the number of scattering paths contributing to the two-photon correlation function Eq.~\eqref{eq_Gamma_r2} becomes exponentially large. We thus must rely on a diagrammatic expansion of the four-field average to identify the scattering sequences that will contribute the most to the 2p-CBS. This requires great care in the reflection geometry since sequences with few scattering events contribute significantly and a rich variety of diagrams may play a prominent role for strong disorder. With the help of two complementary approaches, one based on a Feynman-path type decomposition and the other on a random matrix theory formulation, we show that the correlator in Eq.~\eqref{eq_Gamma_r2} is dominated by the bi-diffuson and bi-cooperon represented in Fig.~1d, in the regime where the transport mean free path of light, $\ell$, exceeds the wavelength $\lambda$ (see Supplementary Section 5 for details). This correlator is expressed as
\begin{equation} \label{eq_Gamma_diff}
    \Gamma_{ba} = \Gamma_0 \left[ 1 + F\left(\vert \mathbf{q}_a - \mathbf{q}_b \vert \right)^2 \right],
\end{equation}
where $\Gamma_0$ is the amplitude of the bi-diffuson and 
$F(\mathbf{q})$ is a lineshape function between 0 and 1, proportional to the transverse Fourier transform of the diffuse intensity. The latter is a decaying function of range $1/\ell$, indicating that most photons experience a few scattering events before being reflected. It is instructive to compare this result with the mean reflection coefficient $R_{ba}=\overline{\vert r_{\mathbf{q}_b,\mathbf{q}_a} \vert^2}$ that characterizes the classical 1p-CBS. Using the same diagrammatic framework and assuming negligible single scattering contribution, we obtain
\begin{equation} \label{eq_R_ba}
R_{ba}=R_0\left[ 1 + F\left(\vert \mathbf{q}_a + \mathbf{q}_b \vert \right)\right],
\end{equation}
where $R_0$ is the amplitude of the diffuson. This means that the contrast of 2p-CBS is simply the square of the contrast of 1p-CBS, in the limit $k\ell \gg 1$ and in the absence of single scattering contribution in the 1p-CBS. For a semi-infinite disordered medium without absorption, $F(q)\simeq1-2q\ell$ in the vicinity of the backscattering angle $q=0$ \cite{RevModPhys.71.313}, so that the 2p-CBS is predicted to be cone shaped, with a width approximately half that of the 1p-CBS cone. Finally, we note that the result in Eq.~\eqref{eq_Gamma_diff} is surprisingly different from the total intensity correlation function $\overline{R_b R_a}$, where $R_a=\sum_{a'}\vert r_{\mathbf{q}_a\mathbf{q}_a'} \vert ^2$, which is known to be dominated by long-range contributions made of pairs of conjugated propagating fields that exchange diffusing partners inside the medium \cite{mello88, 1995_Rogozkin_PRB, 2015_Hsu_PRL}. We show in the Supplementary Section 5 that similar long-range contributions occur in the evaluation of $\Gamma_{ba}$ but have a relative weight $\sim 1/k\ell$ with respect to the bi-diffuson and bi-cooperon.

\begin{figure*}[hbt!]
 \centering
 \includegraphics[scale=0.45]{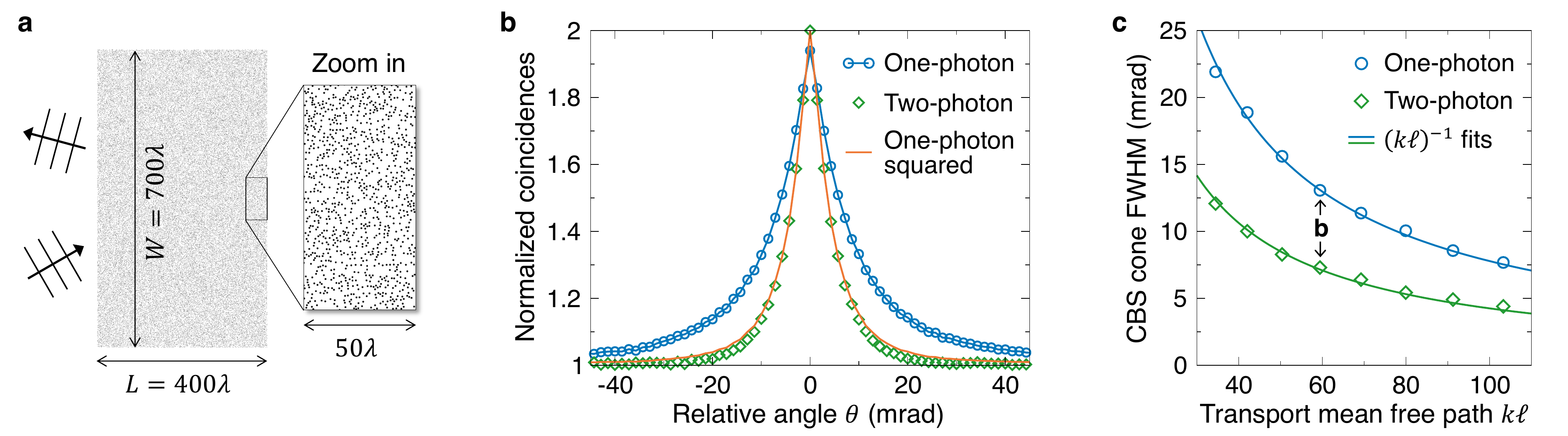}
 \caption{\textbf{Two-photon coherent backscattering from disordered samples in diffusive regime.} \textbf{a}, Relative permittivity profile $\varepsilon _{\rm{r}}(x,y)$ for one realization of disorder. \textbf{b}, Normalized one-photon and two-photon coincidence rates in reflection, $R_{ba}/R_0$ in blue circles, $\Gamma_{ba}/\Gamma_0$ in green diamonds, and $1 + [(R_{ba}/R_0)-0.94]^2$ in orange solid line. The relative angle $\theta$ is as labeled in Fig.~1a,c: $\theta=(q_a+q_b)/k$ for 1p-CBS, $\theta=(q_a-q_b)/k$ for 2p-CBS. \textbf{c}, FWHM of the one-photon and two-photon CBS cones. Symbols are numerical data, and solid lines are $0.78/(k\ell)$ and $0.43/(k\ell)$.}
 \label{fig:4}
\end{figure*}

To rigorously test these predictions, we perform numerical simulations that directly solve the scalar wave equation in two dimensions, $\left[ {\nabla ^2} + k^2\varepsilon _{\rm{r}}(x,y) \right] \psi(x,y) = 0$,  with no approximation beyond spatial discretization.
Evaluating the two-photon correlation function through Eq.~\eqref{eq_Gamma_r2} requires the full $N \times N$ reflection matrix $r$ for all incoming and outgoing states, averaged over a large number of disorder realizations to suppress speckle fluctuations.
The disordered media should have width $W \gtrsim 60\, \ell$ to resolve the 2p-CBS cone shape and thickness $L \gg \ell$ to be in the diffusive regime with sufficient long trajectories that contribute to the sharpness of the cone at small angles \cite{akkermans2007mesoscopic}.
While full-wave reflection matrix computations of such large systems would normally take prohibitive amount of computing resources, some of us recently developed a new scattering-matrix computation method called Schur complement scattering analysis (SCSA) \cite{2022_Lin_arXiv} that is many orders of magnitude more efficient.
Using SCSA, we compute 4,000 distinct reflection matrices in plane-wave basis for different disorder realizations, each consisting of 56,000 randomly positioned $0.8\lambda$-diameter dielectric cylinders in air with 10\% filling fraction (Fig.~\ref{fig:4}a). The transport mean free path is $\ell = 9.5\lambda$ (see Methods for details).

Figure~\ref{fig:4}b shows the numerically calculated 1p-CBS (blue circles) and 2p-CBS (green diamonds) cones; data across the full angular range are shown in Supplementary Fig.~S9. We find the peak-to-background ratio of the 1p-CBS cone to be 1.94; the reduction below 2 comes from single scattering in reflection which does not contribute to the cone \cite{1995_vanTiggelen_EPL, 1996_Amic_JPA}.
We indeed observe a sharp 2p-CBS cone, validating our analytic prediction.
The enhancement factor of the 2p-CBS cone is found to be 2 with no reduction; this is because in the Klyshko’s advanced wave picture for two-photon coincidence, photons traverse the system twice  so they must be scattered at least twice, with no single-scattering contribution left.
The 2p-CBS contrast (green diamonds) is narrower and agrees with the square of the 1p-CBS contrast (orange solid line), in agreement with the analytic prediction above.

To investigate the dependence on the transport mean free path, we vary the filling fraction between 6\% and 17\%. The full set of 1p-CBS and 2p-CBS data are shown in Supplementary Fig.~S10.
Figure~\ref{fig:4}c summarizes the transport mean free path dependence of the numerically computed angular FWHM of the 1p-CBS and 2p-CBS cones, which are well described by single-parameters fits of $0.78/(k\ell)$ for 1p-CBS, $0.43/(k\ell)$ for 2p-CBS. 

The fact that the 2p-CBS cone has a smaller width and larger slope near the cone's peak suggests that it could yield a better estimate of the transport mean free path than 1p-CBS. To make this intuition quantitative, we evaluated the Cramer--Rao lower bound on the parameter $\ell$, which sets a lower bound on the variance of any estimator of $\ell$. It is given by the inverse of the Fisher information matrix $\mathcal{F}$. An important property of 1p-CBS and 2p-CBS is the presence of speckle fluctuations in the CBS map which add up to the Poisson fluctuations of the detector. 
Specifically---and this is a key aspect of the following discussion---the statistics of two-photon speckle built with an EPR state made of a large number of modes $N$ is identical to the statistics of one-photon speckle, in sharp contrast with the two-photon speckle of non-entangled states or mixed states \cite{beenakker2009two}.
Taking into account both the speckle and the Poisson noise, we find the elements of the matrix $\mathcal{F}$ for 2p-CBS to be (Supplementary Section 7)
\begin{equation}
\mathcal{F}^{(2p)}_{ij}= \sum_b \frac{N_r}{N_r+N_{ba}} \frac{\partial_{\theta_i} N_{ba} \,  \partial_{\theta_j}N_{ba}}{N_{ba}},
\label{FisherPoissonSpeckle}
\end{equation}
where $N_{ba} \propto \Gamma_{ba}$ is the number of coincidences detected during the acquisition time, $\theta_i$ are the unknown parameters of the problem at hand, $N_r$ is the number of disorder realizations, and the sum runs over the independent positions (separated by a distance larger than the angular size of a speckle grain) probed by the detector $D_b$. 
Interestingly, in the limit where speckle noise dominates ($N_r \ll N_{ba} $), $\mathcal{F}_{ij}$ becomes independent of the amplitude $\Gamma_0$ of the coincidence rate; in particular, in the situation where only $\ell$ is unknown, we get
$\mathcal{F}^{(2p)}_{\ell \ell}=N_r \sum_b (\partial_\ell \Gamma_{ba}/\Gamma_{ba})^2$. Similarly, for 1p-CBS, we find $\mathcal{F}^{(1p)}_{\ell \ell}=N_r \sum_b (\partial_\ell R_{ba}/R_{ba})^2$. With the theoretical expressions established previously for $\Gamma_{ba}$ and $R_{ba}$ [see Eq.~\eqref{eq_Gamma_diff} and Eq.~\eqref{eq_R_ba}], we immediately conclude that $\mathcal{F}^{(2p)}_{\ell \ell}=4 \mathcal{F}^{(1p)}_{\ell \ell}$ in the vicinity of the cone center. Remarkably, this result holds independently of the precise expression of the CBS profile $F(q)$, because the 2p-CBS contrast is  simply the square of the 1p-CBS contrast. We conclude that the Cramer-Rao lower bound on $\ell$ is reduced by a factor $4$ using 2p-CBS of EPR state instead of 1p-CBS, in the common situation where Poisson noise is negligible (long integration time).  

To summarize, we experimentally observed coherent backscattering of maximally-entangled photon pairs from a dynamically changing scattering medium. We provided an in-depth analysis of the fundamental processes governing the phenomenon, revealing new types of diagrams that determine the scattering process, and which are also absent in classical coherent backscattering. In particular, we find that the two-photon CBS shape is precisely the square of the classical shape in diffusive media, as verified by full-wave numerical simulations. Consequently, the Cramer--Rao lower bound for estimating the transport mean free path from the two-photon CBS shape is four times lower than the bound for classical CBS. The narrower CBS shape can be attributed to the fact that correlations between entangled photons mimic propagation of a single photon at half the wavelength. While such wavelength scaling is typically sensitive to dephasing and noise, we find that in two-photon CBS it prevails scattering and disorder averaging. Finally, we note that since both the Klyshko picture provided to understand two-photon CBS and the cooperon object employ the optical reciprocity principle, it would be interesting to study its role in two-photon CBS by utilizing reciprocity breaking techniques~\cite{muskens2012partial,Bromberg2016}.

\section*{METHODS}
\textit{Experiment---}We use a CW laser at $\lambda_p = 404$ nm (OBIS, Coherent) to pump a $h=2$ mm long periodically-poled potassium titanyl phosphate (PPKTP) crystal in a collinear type-0 degenerate configuration. Pairs of degenerate entangled photons (signal and idler) at $\lambda = 808$ nm are thus generated via spontaneous parametric down conversion (SPDC). The beam waist at the incidence plane of the crystal is approximately $w_p = 550$ $\mu$m, yielding spatially entangled photons with a Schmidt number of  $K = \frac{k_{p}w_{p}^{2}}{4h} \approx 588$ \cite{law2004analysis} . The remainder of the pump beam is deflected via a dichroic mirror located right after the crystal. The down-converted photons are selected via interference filters at $\lambda = 809 \pm 81$ nm. In the 2p-CBS experiment, both photons are imaged via a 4f system ($f=150$ mm) onto a rotating, polarization preserving diffuser of 1 inch diameter (Luminit),  behind which a plane mirror is placed. The measured scattering angle of the diffuser, defined by the $1/e$ width of its disorder-averaged far-field intensity distribution [see Eq.~(3.3) of Supplementary Section 3], was measured to be  $\theta_0 \approx 4.4$ mrad. The reflected light is then scattered once more off the rotating diffuser, and collected by a lens ($f_4=200$ mm) and two fiber-coupled single photon detectors (Excelitas SPCM-AQ4C) of radius $\sigma=50$ $\mu$m located at the Fourier plane of the diffuser. The coincidence circuit is implemented using Swabian Instruments' Time Tagger 20 with a temporal coincidence window of $800$ ps. The figures in this article presenting the coincidence count rates are corrected for the accidental coincidence counts. In the 1p-CBS experiment, the idler photon is detected with a lens ($f_1=150$ mm) and a stationary fiber-coupled detector located at the Fourier plane of the crystal, whereas the signal photon is directed into the circuit mentioned above. The number of modes in our experiment $N$, is smaller than the Schmidt number $K$ by a factor of $\approx 5$, given by the square of the ratio of the SPDC divergence angle ($\approx 20$ mrad) and the scattering angle of the diffuser $\theta_0$. For a more detailed description of the experimental setup, see Supplementary Section 1.

\textit{Numerical simulations---}Numerical simulations are performed by solving the scalar wave equation in two dimensions, $[ {\nabla ^2} + k^2\varepsilon_{\rm{r}}(x,y) ] \psi(x,y) \linebreak[0]  = 0$, using finite-difference discretization with grid resolution $\Delta x = \lambda/10$ and subpixel smoothing~\cite{2006_Farjadpour_OL}\hspace{0.01mm}.
Each disordered medium consists of 56,000 randomly positioned dielectric cylinders with refractive index $n=1.5$ and diameter $0.8\lambda$ in air, inside a region with width $W = 700\lambda$ and varying thickness $L$.
Periodic boundary condition is used in the direction along $W$ to mimic an infinite system, so the angle is discretized at resolution $\delta\theta=\lambda/W$ at small angles.
Perfectly matched layer (PML)~\cite{2005_Gedney_book_chapter}\hspace{0.01mm} is used in the direction along $L$ to implement an outgoing boundary.
Note that since the 2p-CBS cone has an angular FWHM of $0.43/(k\ell)$, a system width of $W > 4\lambda/{\rm FWHM} \approx 58 \ell$ is needed to resolve 5 or more angles with $\delta\theta$ spacing within the half maximum of the 2p-CBS cone.
Here, an individual dielectric cylinder has scattering cross section $\sigma_{\rm sca} = 3.02 \lambda$ and anisotropy factor $g \equiv \langle\cos\theta\rangle = 0.825$, obtained from its differential scattering cross section numerically computed by near-to-far-field transformation for such cylinders at $\Delta x = \lambda/10$.
We obtain the transport mean free path directly through the independent-particle approximation~\cite{carminati_schotland_2021}\hspace{0.01mm} as $l = \left[ \sigma_{\rm sca} \rho (1-g) \right]^{-1}$, where $\rho$ is the number density.
To vary $l$, the thickness $L$ of the scattering region is varied between $L=232\lambda$ and $L=695\lambda$ while fixing the number of dielectric cylinders; the numerically computed average transmission $\bar{T}$ stays between $3.7\%$ and $3.8\%$ for different $L$, in excellent agreement with the analytic prediction of $\bar{T}=\left[ 1 + (2/\pi)(L/l) \right]^{-1}=3.6\%$ (Ref.~[\citenum{RevModPhys.69.731}]), indicating the $l_{\rm t}$ computed from independent-particle approximation is accurate for these configurations.
We use the full-wave solver SCSA~\cite{2022_Lin_arXiv}\hspace{0.01mm} to compute the complete $2N \times 2N$ scattering matrix (which includes the reflection matrices from both sides) without looping over the input states, with $N=1425 \approx 2W/\lambda$ being the number of propagating plane-wave channels on one side.
We perform the simulations for 2,000 realizations of disorder at each $L$, giving 4,000 reflection matrices per thickness.
To further suppress the speckle fluctuations, we also average over 29 vertical slices of matrix $|r_{q_b,q_a}|^2$ and $|(r^2)_{q_b,-q_a}|^2$ centered within $\pm 20$ mrad from $q_a=0$ while excluding $q_b$ at the exact specular direction.
The computations are done on the USC Center for Advanced Research Computing's Discovery cluster. 

\begin{acknowledgments}
\textbf{Funding:} Zuckerman STEM Leadership Program, the Israel Science Foundation (grant No. 2497/21), National Science Foundation (ECCS-2146021), LABEX WIFI (Laboratory of Excellence within the French Program “Investments for the Future”) under references ANR-10-LABX-24 and ANR-10-IDEX-0001-02 PSL*. \textbf{Author contributions:} M.S., O.L. and Y.B. conceived the idea and designed the experiments. M.S. built the experimental setup, performed the measurements and developed the theory for the double-passage configuration. A.G. developed the theory in the multiple scattering regime and Fisher information analysis. H.-C.L. and C.W.H. performed the numerical simulations. All authors analyzed the results and contributed to the manuscript preparation.
\end{acknowledgments}

\newpage

\bibliography{refs}
\end{document}


\preprint{APS/123-QED}

\title{Supplementary Materials}

\maketitle

\section{\label{sec:1} E\lowercase{xperimental setup of quantum} CBS}

\begin{figure}[hbt!]
 \centering
 \includegraphics[scale=0.38]{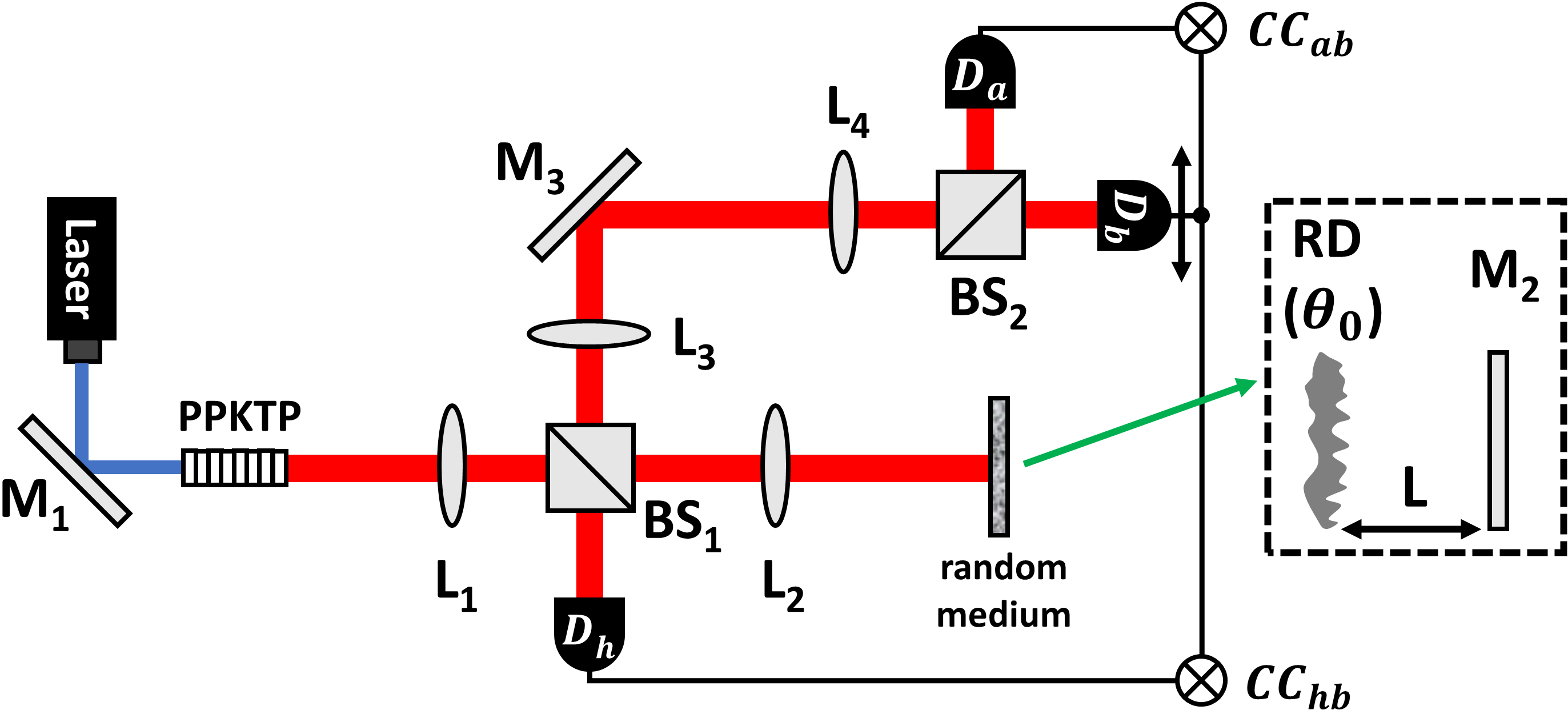}
 \label{fig1}
 \caption{Entangled photon pairs are generated via spontaneous parametric down conversion in a PPKTP crystal. The pairs illuminate the medium, consisting of a diffuser and a mirror placed at a distance L behind it, and the coincidence events of the backscattered photons are collected via two single-photon detectors (static $D_a$ and scanning $D_b$) located at the far-field of the diffuser. For the one-photon experiment, coincidence events of detectors $D_h$ and $D_b$ in the far-field are collected (static $D_h$). M, mirror; L$_{i=1..4}$, lenses; BS, beam-splitter; RD, rotating diffuser; $\theta_0$, scattering angle; D, detector; CC, coincidence counts}
\end{figure}

The experimental setup is depicted in Fig.~S\ref{fig1}. Entangled photon pairs at $\lambda_{SPDC}=808$ nm are generated via spontaneous parametric down conversion (SPDC) by pumping a $h=2$-mm long crystal (PPKTP) with a CW laser at $\lambda_p=404$ nm of waist $w_p = 550~\mu$m. The crystal is then imaged onto a rotating diffuser (RD) using lenses $L_1$ and $L_2$ (focal lengths $f_1=f_2=150$ mm), behind which a plane mirror ($M_2$) is placed. The backscattered photons traverse lenses $L_2$, $L_3$ ($f_2=f_3=150$ mm) and $L_4$ ($f_4=200$ mm), placed in a 6-f configuration to image the far-field plane of the rotating diffuser. Far-field two-photon correlations are measured at the back focal plane of $L_4$, using two single-photon fiber-coupled detectors $D_b$ (transversely scanning) and $D_a$ (stationary) and a coincidence logic ($CC_{ab}$, Swabian TimeTagger 20). The setup is designed to simultaneously measure also the classical (one-photon) CBS shape, using heralded single photons. To this end, we place an additional single photon detector ($D_h$) in the reflection port of BS$_1$, at the far-field of the nonlinear crystal. Detection of a photon by $D_h$ heralds its twin photon in a plane-wave mode that illuminates the sample. The photon is then scattered by the rotating diffuser, and the coincidence counts between the heralding detector $D_h$ and the transversely scanning detector $D_b$ are recorded ($CC_{hb}$). In both the two-photon CBS and one-photon CBS measurements we rotate the diffuser to average over a large number of realizations of  disorder. Moreover, accounting for CBS being a robust interference phenomenon, we use wide interference filters ($\lambda_{filter}=809\pm 81$ nm) to introduce spectral averaging in addition to the disorder averaging. The wide-band filters also help increasing the photon detection rate.

\section{\label{sec:2} C\lowercase{oincidence detection rate}}
\begin{figure}[hbt!]
 \centering
 \includegraphics[scale=0.42]{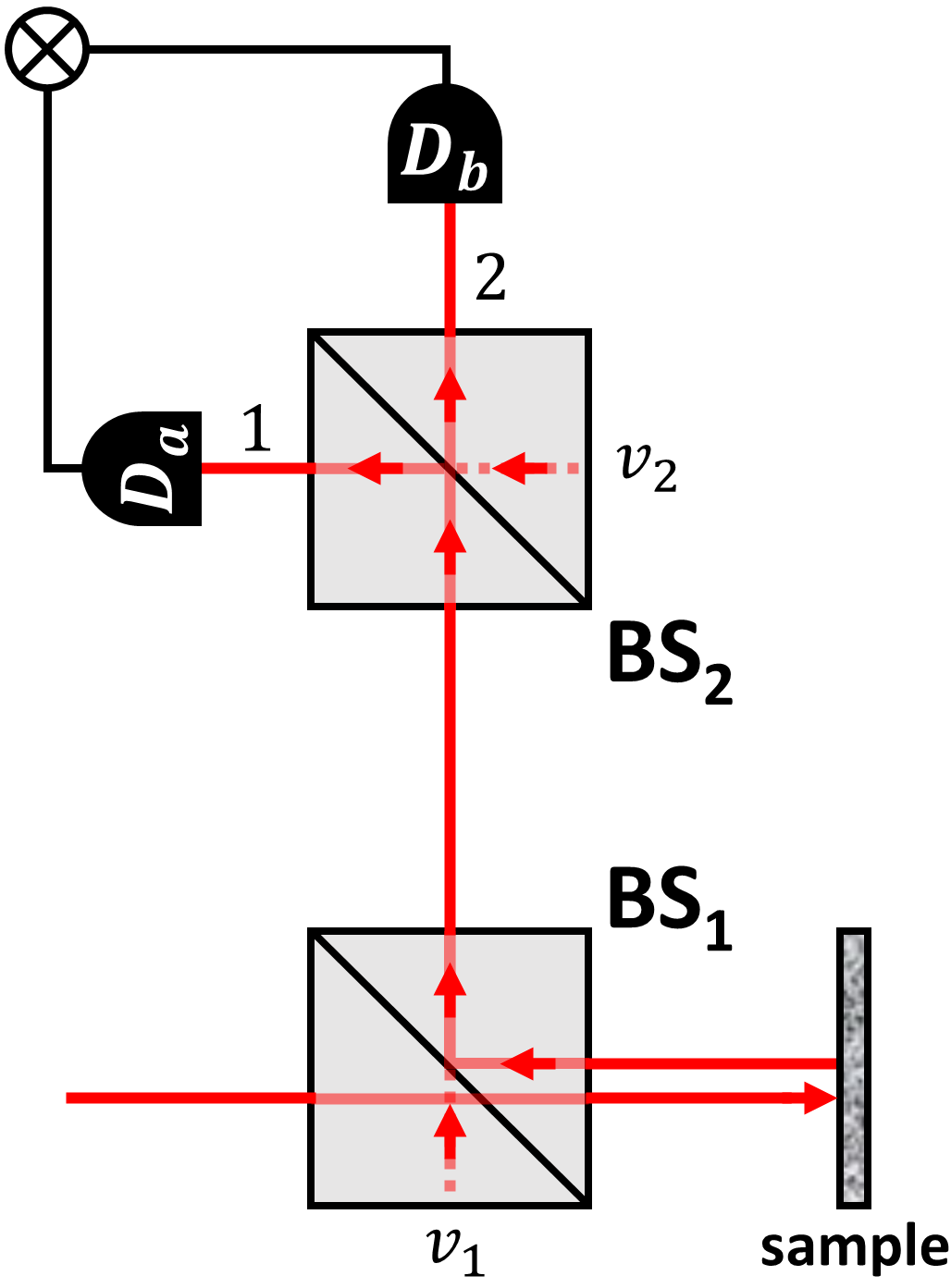}
 \label{fig2}
 \caption{Simplified two-photon CBS experiment. An illustration of the detection process of the backscattered light in ports 1 and 2 after passing through beamsplitters BS$_1$ and BS$_2$. The state for which the expectation value is calculated is $\ket{\psi}\ket{0}_{v_1}\ket{0}_{v_2}$ that describes the photons which backscatter from the sample and the vacuum state from ports $v_1$ and $v_2$.}
\end{figure}

To derive the rate of coincidence events between detectors $D_a$ and $D_b$, $C(\mathbf{q}_a,\mathbf{q}_b)$,  which is proportional to the two-photon correlation function $\Gamma_{ba}$, we express the annihilation operators of the detection modes by the annihilation operators that correspond to modes that backscatter by the sample. Figure~S\ref{fig2} presents a simplified illustration of the detection modes after the backscattered light pass through BS$_1$ and BS$_2$.

We assume lossless symmetric beamsplitters, described by a two-by-two unitary transformation $\left(\begin{array}{cc}
    \frac{1}{\sqrt{2}} & \frac{i}{\sqrt{2}}\\
    \frac{i}{\sqrt{2}} & \frac{1}{\sqrt{2}}
    \end{array}\right)$ relating the input and output modes of the beamsplitter. 
Then, the annihilation operators $\hat{e}_{\mathbf{q},1}$ and $\hat{e}_{\mathbf{q},2}$, describing output modes of BS$_2$ with transverse momentum $\mathbf{q}$, can be expressed using the annihilation operators of the three input modes $\hat{d}_{\mathbf{q},v_1}$, $\hat{d}_{\mathbf{q},v_2}$ and $\hat{d}_{\mathbf{q}}$ by 
\begin{align}
    \hat{e}_{\mathbf{q},1}&=\frac{1}{2}\left(\sqrt{2}\hat{d}_{\mathbf{q},v_2}+i\hat{d}_{\mathbf{q},v_1} - \hat{d}_\mathbf{q}\right) \nonumber \\
    \hat{e}_{\mathbf{q},2}&=\frac{1}{2}\left(i\sqrt{2}\hat{d}_{\mathbf{q},v_2}+\hat{d}_{\mathbf{q},v_1} + i\hat{d}_\mathbf{q}\right).
\end{align}
The disorder-averaged coincidence rate for detecting a photon with transverse momentum $\mathbf{q}_a$ at port 1 and a photon with transverse momentum $\mathbf{q}_b$ at port 2 is given by \vspace{-3.5mm}
\begin{align} \label{Eq:coin_e}
    C(\mathbf{q}_a,\mathbf{q}_b) &= \eta\overline{\langle \hat{e}^\dagger_{\mathbf{q}_b,2}  \hat{e}_{\mathbf{q}_b,2}\hat{e}^\dagger_{\mathbf{q}_a,1} \hat{e}_{\mathbf{q}_a,1}\rangle} \nonumber \\ &=
    \eta\overline{\langle \hat{e}^\dagger_{\mathbf{q}_b,2} \hat{e}^\dagger_{\mathbf{q}_a,1} \hat{e}_{\mathbf{q}_a,1}\hat{e}_{\mathbf{q}_b,2} \rangle},
\end{align}
where in the last step we used $[\hat{e}_{\mathbf{q},1},\hat{e}^\dagger_{\mathbf{q},2}]=0$ to write the coincidence rate operator in normal ordering and $\eta$ accounts for the rate of pairs that illuminate the sample and the overall detection efficiency of the system. The expectation value in Eq.~(\ref{Eq:coin_e}) is computed for the state $\ket{\psi}\ket{0}_{v_1}\ket{0}_{v_2}$ that describes photons that backscatter from the sample, and the vacuum state that couples to port $v_1$ and $v_2$ from the combined system of both beamsplitters. The coincidence rate is thus given by
\begin{align}
    C(\mathbf{q}_a,\mathbf{q}_b) &= \eta \overline{\prescript{}{v_2}{\bra{0}}\prescript{}{v_1}{\bra{0}}\prescript{}{}{\bra{\psi}}\hat{e}^\dagger_{\mathbf{q}_b,2}\hat{e}^\dagger_{\mathbf{q}_a,1}\hat{e}_{\mathbf{q}_a,1}\hat{e}_{\mathbf{q}_b,2}\ket{\psi}_{}\ket{0}_{v_1}\ket{0}_{v_2} \nonumber} \\ &=\eta \overline{|\prescript{}{v_2}{\bra{0}}\prescript{}{v_1}{\bra{0}}\prescript{}{}{\bra{\psi}}\hat{e}_{\mathbf{q}_a,1}\hat{e}_{\mathbf{q}_b,2}\ket{\psi}_{}\ket{0}_{v_1}\ket{0}_{v_2}|^2} \nonumber \\ &= \frac{\eta}{16}\overline{|\bra{0}_{}\hat{d}_{\mathbf{q}_a}\hat{d}_{\mathbf{q}_b}\ket{\psi}_{}|^2}  \nonumber \\ &= \frac{\eta}{16}\overline{\langle :\hat{n}_{\mathbf{q}_b}\hat{n}_{\mathbf{q}_a}:\rangle} =\frac{\eta}{16}\Gamma_{ba}.
\end{align}
We therefore conclude that the rate of coincidence events between two detectors at the output ports of BS$_2$ is given by the two-photon correlation function of the photons that backscatter by the sample.
\vspace{-5mm}
\subsection*{The two-photon correlation function for the EPR state}
To compute the two-photon correlation function of the backscattered photons, we express the input-output relations between the annihilation operators $\hat{c}_{\mathbf{q}_{\alpha'}}$ and $\hat{d}_{\mathbf{q}_{\alpha}}$, representing an input mode with transverse momentum $\mathbf{q}_{\alpha'}$ and an output mode with transverse momentum $\mathbf{q}_{\alpha}$, respectively,
using the reflection matrix $r$ of the sample \cite{beenakker1997random}
\begin{equation} \label{Eq:d}
    \hat{d}_{\mathbf{q}_\alpha}= \sum_{\alpha'} r_{\mathbf{q}_\alpha,\mathbf{q}_{\alpha'}}\hat{c}_{\mathbf{q}_{\alpha'}},
\end{equation}
where the summation is over the $N$ input modes that illuminate the sample.\\
The state of interest in this paper is the EPR state, which in the thin crystal regime can be described as \cite{walborn2010spatial}
\begin{align} \label{Eq:psi}
    \ket{\psi} =\frac{1}{\sqrt{2N}} \sum_{i=1}^{N} \hat{c}_{\mathbf{q}_i}^{\dagger}\hat{c}_{-\mathbf{q}_i}^{\dagger}\ket{0}.
\end{align}
By directly inserting Eq.~(\ref{Eq:d}) and Eq.~(\ref{Eq:psi}) into $\Gamma_{ba}$, one arrives at
\begin{align} \label{Eq:24}
    \Gamma_{ba}&=\frac{1}{2N}\overline{\left|\sum_{a^{\prime},b^{\prime}}\sum_{i}r_{\mathbf{q}_{a},\mathbf{q}_{a^{\prime}}}r_{\mathbf{q}_{b},\mathbf{q}_{b^{\prime}}}\bra{0}\hat{c}_{\mathbf{q}_{a^{\prime}}}\hat{c}_{\mathbf{q}_{b^{\prime}}}\hat{c}_{\mathbf{q}_{i}}^{\dagger}\hat{c}_{-\mathbf{q}_{i}}^{\dagger}\ket{0}\right|^{2}} \nonumber \\
    &= \frac{2}{N}\overline{\left|\sum_{i}r_{\mathbf{q}_{a},-\mathbf{q}_{i}}r_{\mathbf{q}_{b},\mathbf{q}_{i}}\right|^{2}}.
\end{align}

By employing the fact that the reflection matrix $r$ satisfies reciprocity, i.e. $r_\mathbf{q,q^{\prime}}=r_\mathbf{-q^{\prime},-q}$, Eq.~(\ref{Eq:24}) yields
\begin{align} \label{Eq:23}
    \Gamma_{ba}=\frac{2}{N}\overline{\left|\left(r^{2}\right)_{\mathbf{q}_{b},-\mathbf{q}_{a}}\right|^{2}}, 
\end{align}
recovering Eq. (2) of the main text.

\section{T\lowercase{heoretical model for $\Gamma_{ba}$ in the double-passage configuration}}

\begin{figure}[hbt!]
\centering
\includegraphics[width=0.87\linewidth]{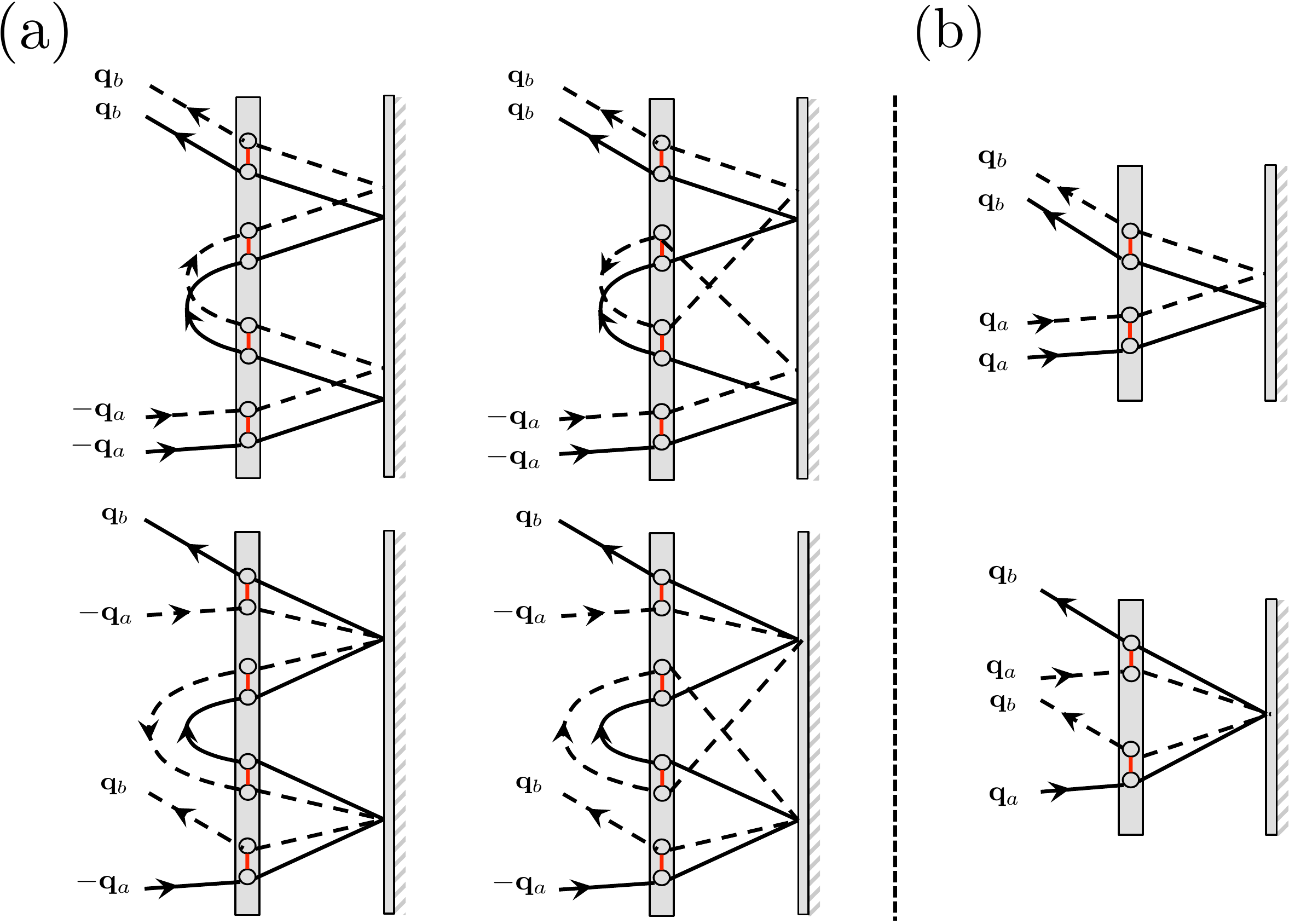}
\caption{Microscopic representation of the double-passage scattering sequences contributing to $\Gamma_{ba}$ (a) and $R_{ba}$ (b). Diffuson and cooperon sequences are shown in the first row and second row, respectively. 
Solid and dashed lines represent the propagating field and its complex conjugate, and open circles stand for the local scattering potentials $V$, which are connected with red lines when they are correlated. In (a), only the four terms that contribute the most to $\Gamma_{ba}$ in the limit $kL\theta_0^2 \gg 1$ are shown.}
\label{FigDiagDouble}
\end{figure}

In the double-passage configuration, a microscopic model for the statistical properties of the reflection matrix can be constructed,
based on the scattering properties of the diffuser and free-space propagation between the medium and the mirror. First, considering the diffuser as a random phase screen, we represent the scattering process by means of a local potential $V$, which obeys Gaussian statistics. The latter is entirely characterized by the real-space correlation function $\overline{ V_{\boldsymbol{\rho},\boldsymbol{\rho}}V^*_{\boldsymbol{\rho}',\boldsymbol{\rho}'}}=\exp\left[-\frac{\left(\boldsymbol{\rho}-\boldsymbol{\rho}^{\prime}\right)^{2}}{(\xi_{0}/\pi)^{2}}\right]$, where $\boldsymbol{\rho}$ and $\boldsymbol{\rho}'$ label transverse positions inside the thin disordered screen  and $\xi_0=\lambda/\theta_0$ is the characteristic coherence length of the diffuser, set by the scattering angle  $\theta_0$ and the wavelength $\lambda$. Second, due to the finite spacing $L$ between the diffuser and the mirror, one must take into account the free space propagation between two scattering events.  This is done by introducing the Fresnel matrix $H$ which is diagonal in momentum space: $H_{\mathbf{q},\mathbf{q}'}\triangleq H_{\mathbf{q}}\delta_{\mathbf{q},\mathbf{q}'} = \exp\left(-i\frac{{q^2}}{2k}d\right)\delta_{\mathbf{q},\mathbf{q}'}$, where $d=2L$. In this way, the reflection matrix takes the form
\be
r=VHV,
\label{EqVHV}
\ee
and the disorder-averaged two-photon correlation function $\Gamma_{ba}$ reads 
\begin{widetext}
\begin{align}
\Gamma_{ba} & \propto \overline{\left|(r^2)_{\mathbf{q}_b,-\mathbf{q}_a}\right|^2}=\sum_{a^{\prime},b^{\prime}}\overline{r_{\mathbf{q}_b,\mathbf{q}_{a^{\prime}}}r_{\mathbf{q}_{a^{\prime}},-\mathbf{q}_{a}}r_{\mathbf{q}_b,\mathbf{q}_{b^{\prime}}}^{*}r_{\mathbf{q}_{b^{\prime}},-\mathbf{q}_{a}}^{*}}\nonumber \\
 & =\sum_{\substack{a^{\prime},m,n\\
b^{\prime},m^{\prime},n^{\prime}
}
}H_{\mathbf{q}_m}H_{\mathbf{q}_{m^{\prime}}}^{*}H_{\mathbf{q}_n}H_{\mathbf{q}_{n^{\prime}}}^{*}\overline{V_{\mathbf{q}_b,\mathbf{q}_m}V_{\mathbf{q}_m,\mathbf{q}_{a^{\prime}}}V_{{\mathbf{q}_{a^{\prime}},\mathbf{q}_n}}V_{\mathbf{q}_n,-\mathbf{q}_a}V_{{\mathbf{q}_{b},\mathbf{q}_{m^{\prime}}}}^{*}V_{{\mathbf{q}_{m^{\prime}},\mathbf{q}_{b^{\prime}}}}^{*}V_{{\mathbf{q}_{b^{\prime}},\mathbf{q}_{n^{\prime}}}}^{*}V_{\mathbf{q}_{n^{\prime}},-\mathbf{q}_{a}}^{*}}.
\label{eq:1}
\end{align}
\end{widetext}
By employing the complex Gaussian moment theorem \cite{goodman2015statistical}, one can split Eq.~(\ref{eq:1}) into pairwise contractions leading to a total of $4!=24$ terms. Each pair contraction is of the form
\begin{equation} 
\overline{V_{\mathbf{q}_{\alpha},\mathbf{q}_{\beta}}V_{\mathbf{q}_{\gamma},\mathbf{q}_{\delta}}^{*}}\propto
\exp\left[-\frac{\left(\mathbf{q}_{\alpha}-\mathbf{q}_{\beta}\right)^{2}}{k^2\theta_0^2}\right]
\delta_{\mathbf{q}_{\beta}-\mathbf{q}_{\delta},\mathbf{q}_{\alpha}-\mathbf{q}_{\gamma}}.
\label{1223}
\end{equation}
The term $\delta_{\mathbf{q}_{\beta}-\mathbf{q}_{\delta},\mathbf{q}_{\alpha}-\mathbf{q}_{\gamma}}$ in Eq.~(\ref{1223}) is a signature of the memory effect of thin diffusers \cite{freund1988memory,feng1988correlations}: if an incident beam with $\mathbf{q}_\beta$ is tilted to $\mathbf{q}_\delta$, the output beam also gets tilted by the same amount from $\mathbf{q}_\alpha$ to $\mathbf{q}_\gamma$. On the other hand, the Gaussian prefactor in Eq.~\eqref{1223} describes the anisotropic scattering of the phase screen. In order to find the leading contributions among the $24$ terms of Eq.~\eqref{eq:1}, we can first
analyze the case $\mathbf{q}_{a}=\mathbf{q}_{b}=\mathbf{0}$. After a lengthy but straightforward calculation, we find that
\begin{widetext}
\be
\Gamma_{ba}(\mathbf{q}_{b}=\mathbf{0}, \mathbf{q}_{a}=\mathbf{0}) \propto 2\left[1+\frac{1}{\sqrt{1+\left(kL\theta_{0}^{2}\right)^{2}/2}}+\frac{1}{\sqrt{1+\left(kL\theta_{0}^{2}\right)^{2}/4}}\right].
\label{1224}
\ee
\end{widetext}
In the limit $kL\theta_{0}^{2}\gg1$, the last two terms in Eq.~\eqref{1224} become negligible. This limit is met in the experiment, as the expansion of the diffracted beam, $L\theta_0$,  exceeds  the coherence length of the diffuser $\xi_0$. Thus, to study the lineshape of the two-photon CBS, it is enough to consider the four leading terms which are responsible for the first term in Eq.~\eqref{1224} in the direction $\mathbf{q}_{a}=\mathbf{q}_{b}=\mathbf{0}$. They are represented in Fig.~\ref{FigDiagDouble}(a). Evaluation of these terms, which are twice degenerate, yields
\begin{multline}\label{111} 
\Gamma_{ba}\propto\left\{ 1+\exp\left[-\frac{\left(2L\theta_{0}\right)^{2}}{2}\left(\mathbf{q}_{a}-\mathbf{q}_{b}\right)^{2}\right]\right\} \\
\times\exp\left[-\frac{\left(\mathbf{q}_{a}+\mathbf{q}_{b}\right)^{2}}{4k^{2}\theta_{0}^{2}}\right], 
\end{multline}
recovering Eq.~(3) of the main text. This equation can also be expressed in terms of angular coordinates $\boldsymbol{\theta}=\mathbf{q}/k$ as
\begin{multline} \label{123}
    \Gamma_{ba} \propto \left\{ 1+\exp\left[-\frac{\left(2kL\theta_{0}\right)^{2}}{2}\left(\boldsymbol{\theta}_{a}-\boldsymbol{\theta}_{b}\right)^{2}\right]\right\}\\
    \times\exp\left[-\frac{\left(\boldsymbol{\theta}_{a}+\boldsymbol{\theta}_{b}\right)^{2}}{4\theta_{0}^{2}}\right].
\end{multline}
To account for the finite angular resolution of our setup, we convolve Eq.~(\ref{123}) with the mode profile of the detection apparatus, taken to be Gaussians $I_{a}\left(\boldsymbol{\theta}_{a}\right)$ and $I_{b}\left(\boldsymbol{\theta}_{b}\right)$ of angular width $\delta_a=\delta_b=\sigma/f_{4}$ (see Section \ref{sec:1}), where $\sigma=50$ $\mu$m is the radius of the collecting fibers, 
\begin{equation}
    \Gamma_{ba}^{\text{conv}}\propto \Gamma_{ba} \ast I_{a}\left(\boldsymbol{\theta}_{a}\right) \ast I_{b}\left(\boldsymbol{\theta}_{b}\right).
\end{equation}
Performing the convolution yields a CBS width of $\theta_{2p} = \sqrt{\delta_{2p}^{2} + (2kL\theta_0)^{-2}}$, where $\delta^2_{2p} \triangleq \delta_a^2+\delta_b^2 = 2\sigma^2/f_4^2$.

It is also instructive to evaluate the 1p-CBS with the model in Eq.~\eqref{EqVHV}.  The mean reflection coefficient $R_{ba}$ is given by
\begin{align}
R_{ba} & = \overline{\left|r_{\mathbf{q}_b,\mathbf{q}_a}\right|^{2}} \nonumber \\
&=\sum_{m,m^{\prime}}H_{\mathbf{q}_m}H_{\mathbf{q}_{m^{\prime}}}^{*}\overline{V_{\mathbf{q}_b,\mathbf{q}_m}V_{\mathbf{q}_m,\mathbf{q}_a}V_{\mathbf{b}_m,\mathbf{q}_{m^{\prime}}}^{*}V_{\mathbf{b}_{m^{\prime}},\mathbf{q}_a}^{*}},
\end{align}
where the average can be contracted into 2 terms, represented in Fig.~\ref{FigDiagDouble}(b). Using the correlator~(\ref{1223}), we find
\begin{multline}\label{1122}
    R_{ba} \propto \left\{ 1+\exp\left[-\frac{\left(L\theta_{0}\right)^{2}}{2}\left(\mathbf{q}_{a}+\mathbf{q}_{b}\right)^{2}\right]\right\}\\
    \times\exp\left[-\frac{\left(\mathbf{q}_{a}-\mathbf{q}_{b}\right)^{2}}{2k^{2}\theta_{0}^{2}}\right],  
\end{multline}
which is Eq. (4) of the main text. This equation can also be expressed in terms of angular coordinates $\boldsymbol{\theta}=\mathbf{q}/k$ as
\begin{multline} \label{121233}
    R_{ba} \propto \left\{ 1+\exp\left[-\frac{\left(kL\theta_{0}\right)^{2}}{2}\left(\boldsymbol{\theta}_{a}+\boldsymbol{\theta}_{b}\right)^{2}\right]\right\}\\
    \times\exp\left[-\frac{\left(\boldsymbol{\theta}_{a}-\boldsymbol{\theta}_{b}\right)^{2}}{2\theta_{0}^{2}}\right].
\end{multline}
Here, we should account for the angular resolution of detector $D_b$ and the finite angular divergence of the heralding photon, determined by the angular resolution of the heralding detector that is taken to be a Gaussian, $I_{h}\left(\boldsymbol{\theta}_{a}\right)$, of width $\delta_{h} = \sigma/f_{1}$ (see Section \ref{sec:1}). Thus
\begin{equation} \label{1233}
    R_{ba}^{\text{conv}}\propto R_{ba} \ast I_{h}\left(\boldsymbol{\theta}_{a}\right) \ast I_{b}\left(\boldsymbol{\theta}_{b}\right).
\end{equation}
Performing the convolution while assuming $\delta_h \backsimeq \delta_b$ yields a CBS width of $\theta_{1p} = \sqrt{\delta_{1p}^{2} + (kL\theta_0)^{-2}}$, where $\delta^2_{1p} \triangleq \delta_h^2+\delta_b^2$.
\vspace{-3mm}
\section{F\lowercase{its to the experimental data}}
\begin{figure}[hbt!]
 \centering
 \includegraphics[scale=0.29]{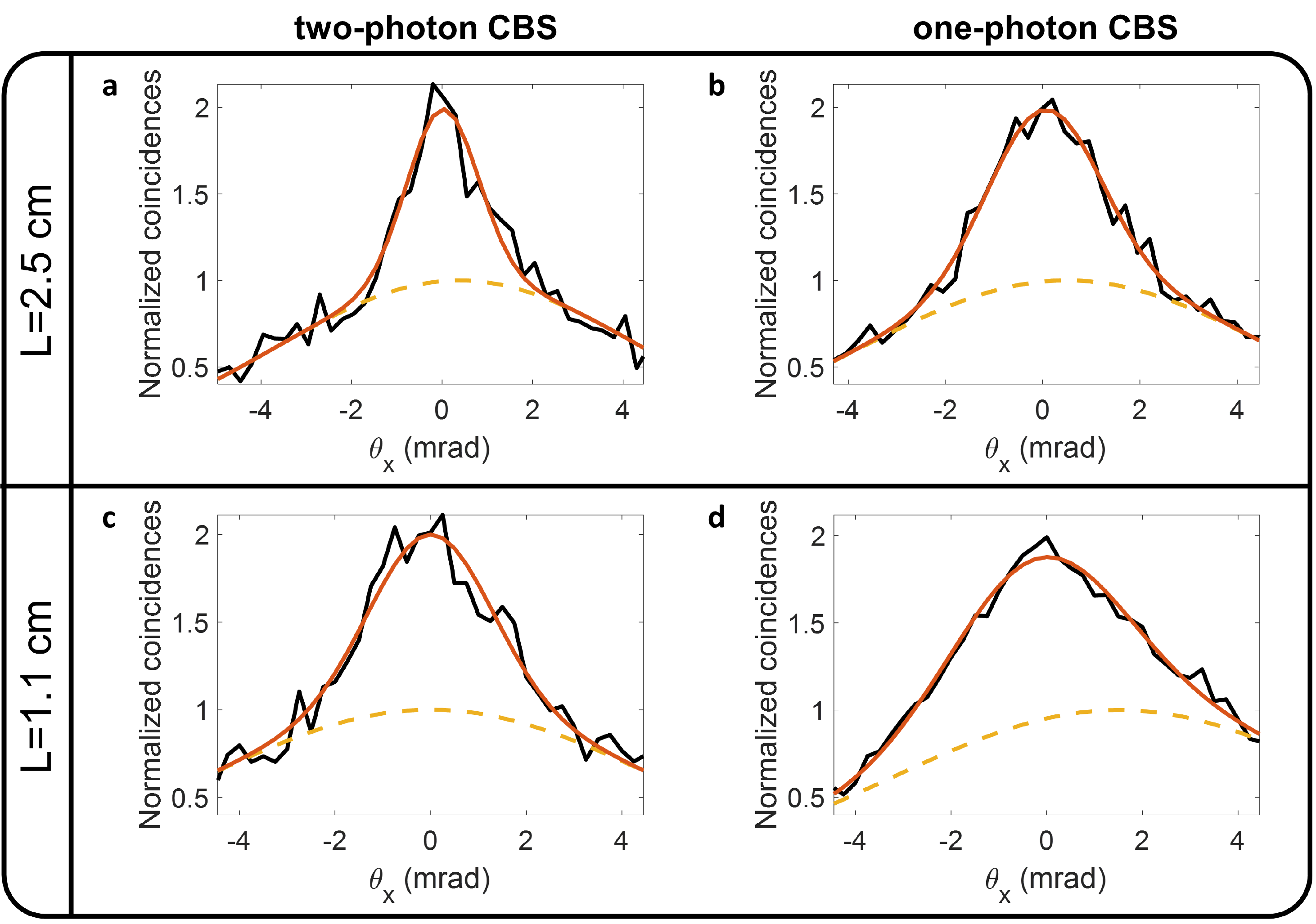}
 \label{fig:data_fits}
 \caption{Two-photon and one-photon CBS data and fits for two diffuser-mirror spacings $L=2.5$ cm an $L=1.1$ cm. Orange line is the fit to the overall expression in each case, and the dashed yellow line is the background fit only.}
\end{figure}

Figure~S4 shows the experimental data of the two-photon CBS [Fig.~S\ref{fig:data_fits}(a)\&(c)] and one-photon CBS [Fig.~S\ref{fig:data_fits}(b)\&(d)] for diffuser-mirror spacings of $L=2.5$ cm (same data in Fig.~2 of the main text) and $L=1.1$ cm. The two-photon CBS and one-photon CBS data were fit to Eq.~\eqref{123} and Eq.~\eqref{121233}, respectively. These fits were used to determine the 2p-CBS and 1p-CBS widths reported in Fig.~3 of the main text. Each of the fits had four fitting parameters accounting for the widths of each Gaussian (background and CBS), as well as the position of the static detector (which could introduce a relative skew between the background and CBS peak). The fourth fitting parameter was taken to account for a global angular shift in the position of the background due to slight misalignment of the mirror behind the diffuser.

\vspace{-4mm}
\section{T\lowercase{heoretical models for $\Gamma_{ba}$ in the multiple scattering regime}}
\label{SecTheory}
Our goal is to evaluate the correlator
\begin{align}
\Gamma_{ba}&=\overline{\vert (r^2)_{\vec{q}_{b}, -\vec{q}_{a}}\vert^2}
\nonumber
\\
&=\sum_{\vec{q}_{a'}, \vec{q}_{b'}} \overline{r_{\vec{q}_{b},\vec{q}_{a'}}r_{\vec{q}_{a'},-\vec{q}_{a}} r_{\vec{q}_{b},\vec{q}_{b'}}^*r_{\vec{q}_{b'},-\vec{q}_{a}}^*}
\label{EqDefGamma}
\end{align}
in the multiple scattering regime, where the simple representation~\eqref{EqVHV} does not hold. For convenience, we absorbed the $2/N$ prefactor appearing in Eq.~\eqref{Eq:23} in the definition of $\Gamma_{ba}$.

Two strategies can be be adopted. In a microscopic treatment, we decompose each reflected field onto all the possible scattering paths and search for the sequences that give non-negligible contributions to the four-field\\
\\
average appearing in Eq.~\eqref{EqDefGamma}. This approach has the advantage to immediately provide a simple microscopic interpretation of the 2p-CBS. However in this framework, it is difficult to properly account for all microscopic diagrams relevant to the reflection geometry. Alternatively, we can perform a singular value decomposition of each reflection matrix in terms of random unitary matrices, and perform an average in the circular unitary ensemble. This second approach has the clear benefit to properly account for all possible contributions to $\Gamma_{ba}$, but misses simple microscopic interpretation \textit{a priori}. A comparison with the path-decomposition framework will make possible to find microscopic interpretation for all leading contributions \textit{a posteriori}.
\vspace{-5mm}
\subsection{Microscopic approach}

Let us first consider Gaussian contributions, which are obtained by pairing reflection coefficients to form averages of complex conjugate pairs:
\begin{widetext}
\be
\Gamma_{ba}\simeq \sum_{\vec{q}_{a'}, \vec{q}_{b'}}
 \overline{r_{\vec{q}_{b},\vec{q}_{a'}} r_{\vec{q}_{b},\vec{q}_{b'}}^*}
 \;
  \overline{r_{\vec{q}_{a'},-\vec{q}_{a}} r_{\vec{q}_{b'},-\vec{q}_{a}}^*}
+   \overline{r_{\vec{q}_{b},\vec{q}_{a'}} r_{\vec{q}_{b'},-\vec{q}_{a}}^*}
\;
    \overline{r_{\vec{q}_{a'},-\vec{q}_{a}} r_{\vec{q}_{b},\vec{q}_{b'}}^*}.
\ee
\end{widetext}
In addition, in the reflection geometry, we must take into account the diffuson (D) and cooperon (C) diagrams for each pair of complex conjugated fields. In a diffuson, the two fields visit the same scatterers in the same order, whereas in a cooperon, the scatterers are visited in the opposite order. Putting aside single scattering, each diffuson sequence possesses a cooperon counterpart. In the following, we ignore for simplicity the special case of single scattering (which can be filtered out by considering cross-polarized channels), and use reciprocity to write
\begin{align}
 \overline{r_{\vec{q}_{\alpha},\vec{q}_{\beta}} r_{\vec{q}_{\gamma},\vec{q}_{\delta}}^*}&=\overline{r_{\vec{q}_\alpha \vec{q}_\beta}r^*_{\vec{q}_\gamma \vec{q}_\delta}}^{(D)} + \overline{r_{\vec{q}_\alpha \vec{q}_\beta}r^*_{\vec{q}_\gamma \vec{q}_\delta}}^{(C)}
\nonumber
\\
&= \overline{r_{\vec{q}_\alpha \vec{q}_\beta}r^*_{\vec{q}_\gamma \vec{q}_\delta}}^{(D)} + \overline{r_{\vec{q}_\alpha \vec{q}_\beta}r^*_{-\vec{q}_\delta -\vec{q}_\gamma}}^{(D)}.
\label{Corr}
\end{align}

In the multiple scattering regime, the field-field correlation function takes the form
\be
\overline{r_{\vec{q}_\alpha \vec{q}_\beta}r^*_{\vec{q}_\gamma \vec{q}_\delta}}^{(D)}=\delta_{\vec{q}_\beta-\vec{q}_\delta, \vec{q}_\alpha-\vec{q}_\gamma}F(\vert \vec{q}_\beta-\vec{q}_\delta \vert),
\label{EqDiffuson}
\ee
where $F(q)$ is the Fourier transform of the reflected intensity profile. It is a decaying function of range $1/\ell$ because most light experiences a few scattering events before being reflected. Close to normal incidence, the latter reads~\cite{gorodnichev90}
\be
F(q) \propto \exp\left[-\frac{2}{\pi} \int_0^{\pi/2}d\beta \, \text{ln}[1-f(\text{tan}^2\beta+q^2\ell^2)],\right]
\label{EqFctCBS}
\ee
where $f(x)=\text{arctan}(\sqrt{x})/\sqrt{x}$ in 3D and $f(x)=1/\sqrt{x+1}$ in 2D. In the limit $q\to 0$, it becomes\vspace{-1mm}
\be
F(q) \propto \frac{1}{(1+ q \ell)^2}. 
\label{EqFLowQ}
\ee
Using the functional form~\eqref{EqDiffuson}, Eq.~\eqref{Corr} reduces to 
\be
 \overline{r_{\vec{q}_{\alpha},\vec{q}_{\beta}} r_{\vec{q}_{\gamma},\vec{q}_{\delta}}^*}=\delta_{\vec{q}_\beta-\vec{q}_\delta, \vec{q}_\alpha-\vec{q}_\gamma}\left[ F(\vert \vec{q}_\beta-\vec{q}_\delta \vert) + F( \vert\vec{q}_\beta+\vec{q}_\gamma \vert)
\right].
\label{CorrDiff}
\ee 
In particular, this identity gives the standard coherent backscattering profile:
\be
\overline{\vert r_{ba} \vert^2}=F(0)+F(\vert \vec{q}_a+\vec{q}_b \vert),
\label{standardCBS}
\ee
where $\vert \vec{q}_a+\vec{q}_b \vert=2k\text{sin}\theta/2$, with $\theta$ the angle between the input and output waves. Upon renormalizing  F(q) by F(0), we recover Eq.~(6) of the main text.  

\begin{figure}[hbt!]
\centering
\includegraphics[width=1\linewidth]{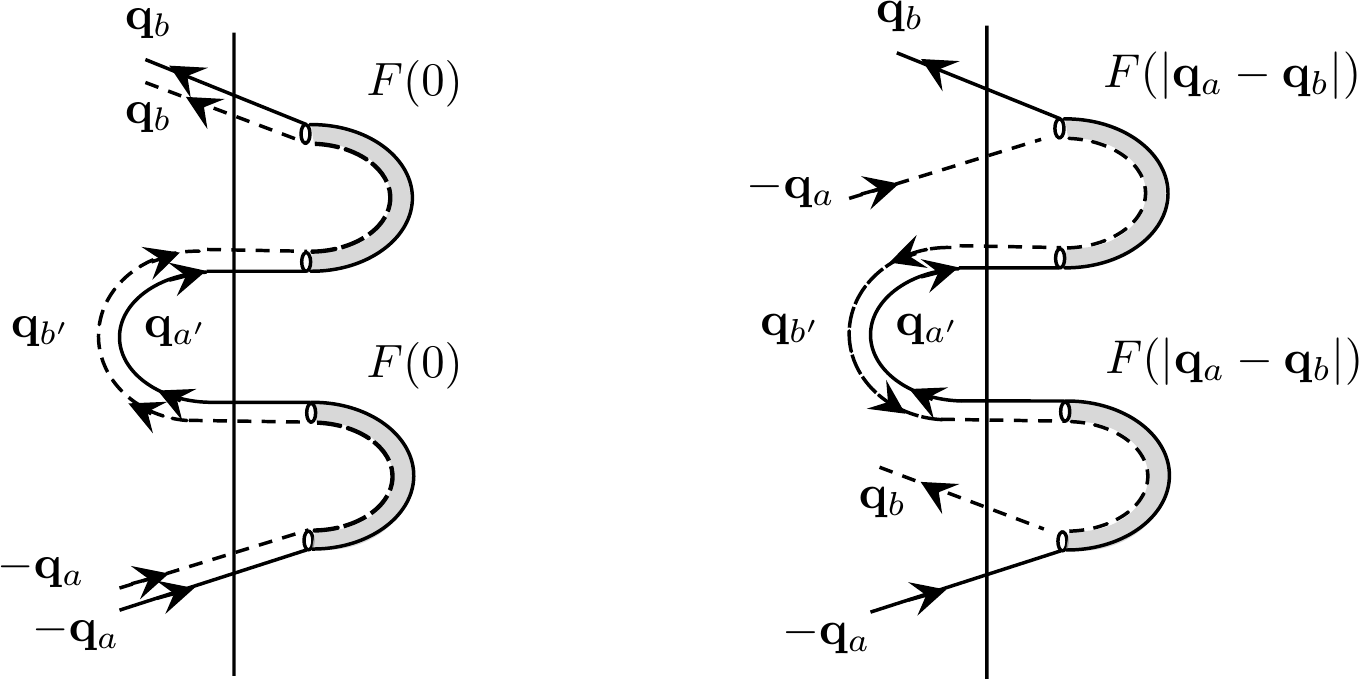}
\caption{Bi-diffuson and bi-cooperon contributions to $\Gamma_{ba}$. Solid lines represent averaged fields, dashed lines their complex conjugates, and shaded tubes diffusive paths; open circles stand for scatterers located at the beginning and the end of multiple scattering sequences.}
\label{FigDiag1}
\end{figure} 

\begin{figure}[hbt!]
\centering
\includegraphics[width=1\linewidth]{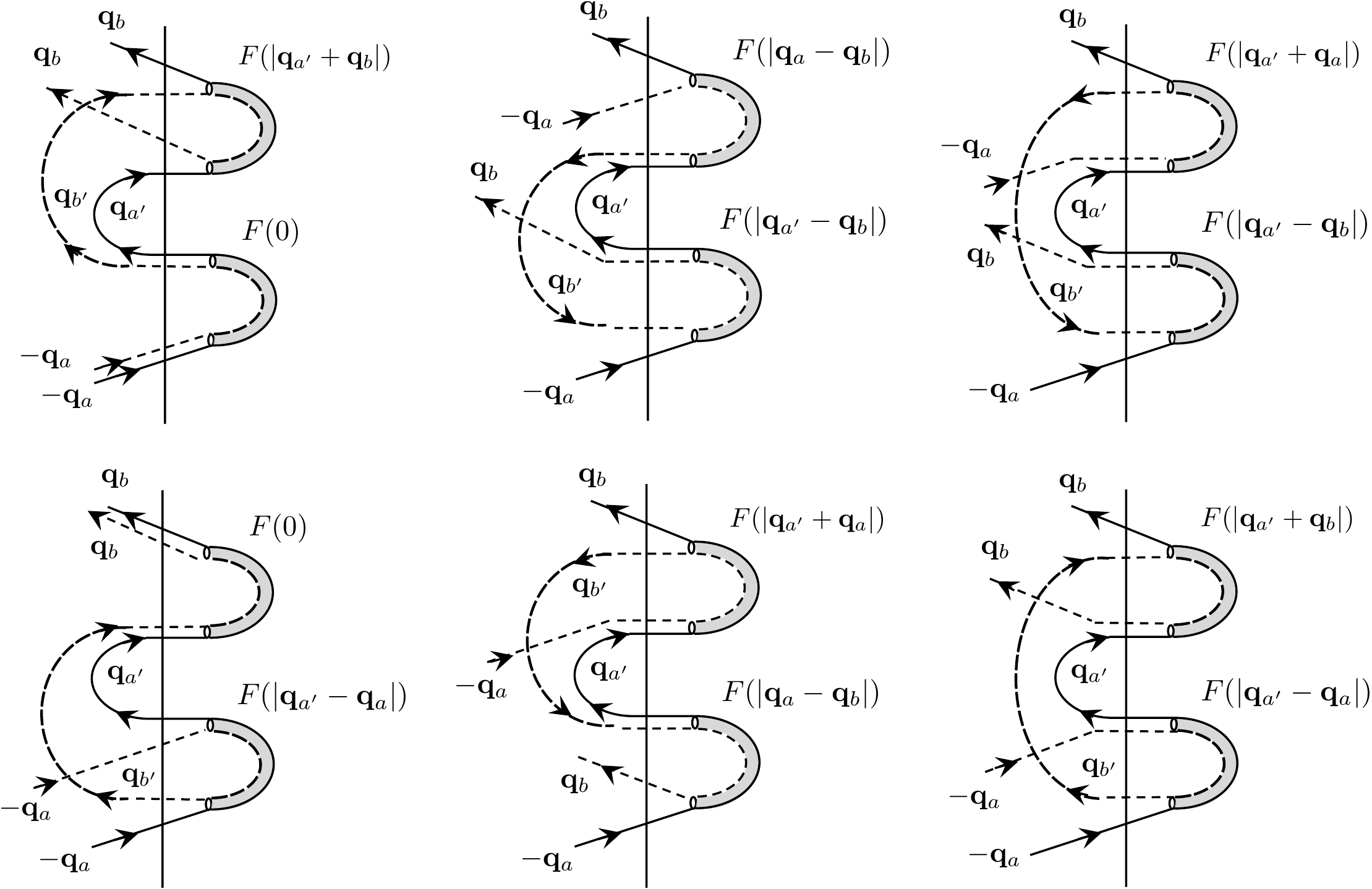}
\caption{Sub-leading Gaussian diagrams contributing to $\Gamma_{ba}$. Diagrams on top of each others have equal weight. In the RMT approach, most of these diagrams are compensated by non-Gaussian contributions.}
\label{FigDiag2}
\end{figure} 

With the help of Eq.~\eqref{CorrDiff}, the normalized coincidence rate takes the form 
\begin{widetext}
\begin{align}
\Gamma_{ba}&=\sum_{\vec{q}_{a'}}\left[ F(0) + F( \vert \vec{q}_{a'}+\vec{q}_{b} \vert) \right] \left[ F(0) + F(\vert\vec{q}_{a'}-\vec{q}_{a}\vert) \right]
\nonumber
\\
 & \;\;\;\;\;\;\;\;\;\;+\left[ F(\vert\vec{q}_{a'}+\vec{q}_{a}\vert)
+ F(\vert-\vec{q}_{a}+\vec{q}_{b}\vert) \right]\left[ F(\vert\vec{q}_{a'}-\vec{q}_{b}\vert) + F(\vert-\vec{q}_{a}+\vec{q}_{b}\vert) \right]
\nonumber
\\
&= N\left[ F(0)^2 + F(\vert\vec{q}_{a}-\vec{q}_{b}\vert)^2 \right] + 2  \sum_{\vec{q}_{a'}} F(q_{a'}) \left[ F(0) + F(\vert\vec{q}_{a}-\vec{q}_{b}\vert) +F(\vert\vec{q}_{a}+\vec{q}_{b}-\vec{q}_{a'}\vert) \right],
\label{EqSolGammaC1}
\end{align}
\end{widetext}
where $N$ is the number of transverse modes probed in the experiment. Each of the eight terms of Eq.~\eqref{EqSolGammaC1} allows for simple microscopic representation. The first two terms correspond to the bi-diffuson and bi-cooperon represented in Fig.~\ref{FigDiag1}, whereas the six others correspond to different combinations of diffusons and cooperons, as shown in Fig.~\ref{FigDiag2}.  The explicit evaluation of all terms using the simple model~\eqref{EqFLowQ} in a slab of finite transverse dimension shows that the first two terms are of order $N$, whereas the other terms are of order $N/k\ell$. Hence, in the limit of weak scattering, $k\ell \gg1$, we get
\be
\Gamma_{ba} \propto F(0)^2 + F(\vert \vec{q}_{a}-\vec{q}_{b} \vert)^2,
\label{EqSolGammaC1Approx}
\ee
which is Eq.~(5) of the main text, up to a renormalization of $F(q)$ by $F(0)$. 

At this stage, it is important to ask whether the Gaussian contributions considered in Eq.~\eqref{EqSolGammaC1Approx}  are the most contributing scattering sequences to $\Gamma_{ba}$. Non-Gaussian contributions may also play an important role,
as is  the case for the total intensity correlation function  $\overline{R_b R_a}$, where $R_a=\sum_{a'}\vert r_{\mathbf{q}_a\mathbf{q}_a'} \vert ^2$. The latter is known to be dominated by long-range non-Gaussian contributions (commonly termed $C_2$ contributions), made of pairs of conjugated propagating fields which exchange diffusing partners inside the medium through a Hikami box. Furthermore, we can wonder whether sub-leading Gaussian terms identified in Eq.~\eqref{EqSolGammaC1} could be compensated by four-field scattering sequences that differ from simple diffusons and cooperons. 

In the following, instead of looking blindly for leading diagrammatic non-Gaussian contributions in reflection, we will adopt a random matrix theory approach, which has the advantage to systematically capture all contributions of the same order $N$.  

\subsection{Random matrix theory approach}

To simplify the RMT treatment, we assume that the disordered system is placed in a waveguide with perfectly reflecting walls. This allows us to identify the directions $\vec{q}$ and $-\vec{q}$ and write Eq.~\eqref{EqDefGamma} as
\begin{align}
\Gamma_{ba}&=\overline{\vert (r^2)_{b a}\vert^2}
\nonumber
\\
&=\sum_{a',b'} \overline{r_{ba'}r_{a'a} r_{bb'}^*r_{b'a}^*}.
\label{EqDefGammaRMT}
\end{align}
\begin{figure*}[t]
\centering
\includegraphics[width=0.9\linewidth]{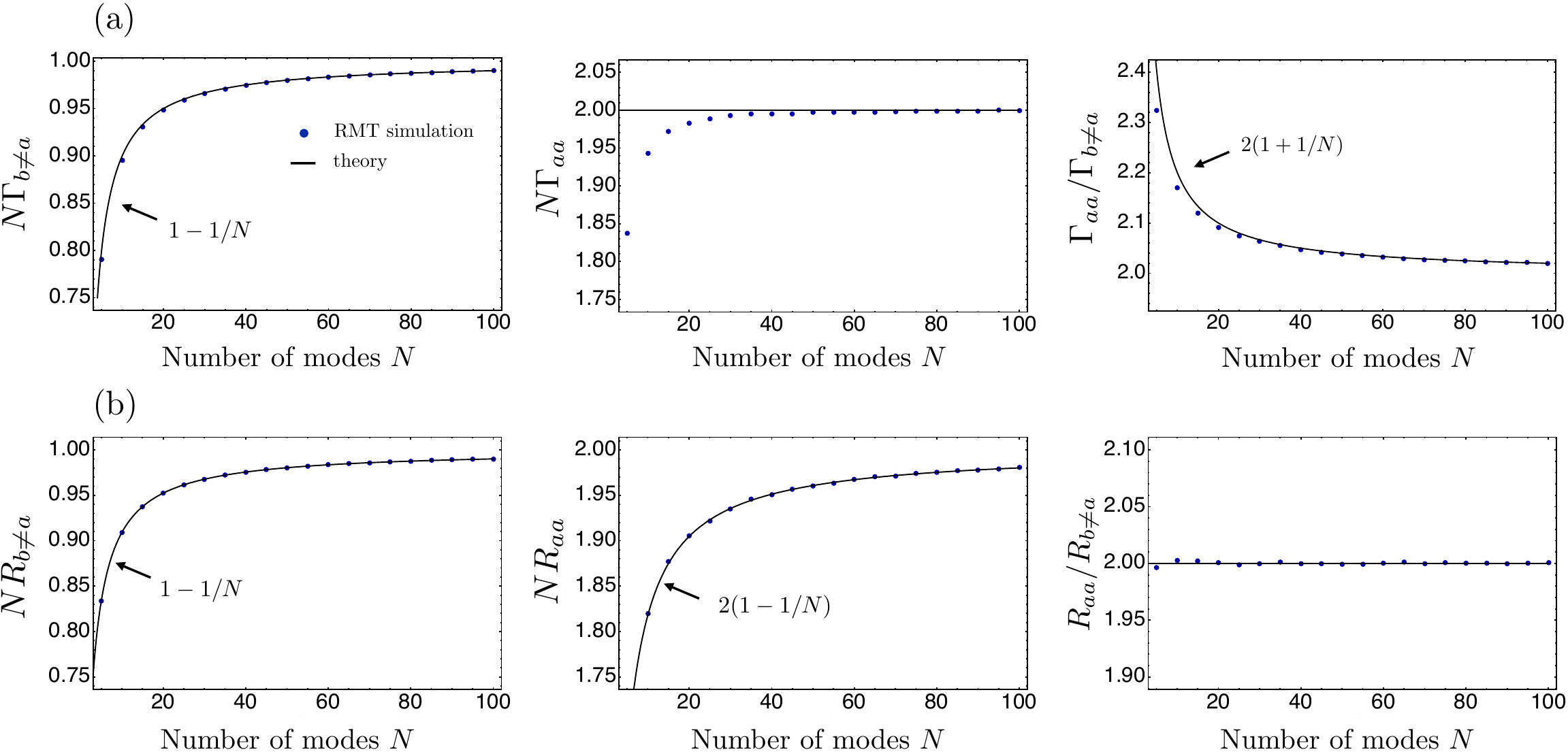}
\caption{ RMT predictions for 2p-CBS (a) and 1p-CBS (b) as function of the number of modes $N$. RMT simulations have been performed by sampling $U$ in the circular unitary group, with $10^5$ configurations for each value of $N$. }
\label{FigRMT}
\end{figure*} 
Furthermore, for a non-absorbing system which preserves time-reversal symmetry, the singular value decomposition of the reflection matrix $r$ can be written as 
\be
r=U^*\sqrt{R}U^\dagger,
\label{EqRSVD}
\ee
where $U$ is a $N\times N$ random unitary matrix and $R$ is a diagonal matrix whose elements are the reflection eigenvalues ($0 \le R_n \le 1$). 
With the decomposition $r_{a'a}=\sum_n \sqrt{R_n}U_{a'n}^*U_{an}^*$ inserted into Eq.~\eqref{EqDefGammaRMT}, the coincidence rate becomes
\begin{widetext}
\be
\Gamma_{ba} = \sum_{n, n', p, p'} \overline{(R_nR_{n'}R_pR_{p'})^{1/2}} \sum_{a', b'}\overline{U_{bn}^*U_{a'n}^*U_{a'n'}^*U_{an'}^*U_{bp}U_{b'p}U_{b'p'}U_{ap'}},
\label{EqGammaRMTDevelopped}
\ee
\end{widetext}
where we split the ensemble average into an average over the unitary matrices and an average over the reflection eigenvalues. In addition, we assume that the unitary matrices are uniformly distributed in the unitary group. This assumption is known to be valid for waveguide of width $W < \ell, L$~\cite{beenakker1997random}. As a result, it cannot capture the precise shape of the 1p-CBS or 2p-CBS, but it presents the great advantage to rigorously access the relative weight of all possible scattering contributions. In the circular unitary ensemble, the average of four doublets $\{U, U^*\}$ as in Eq.~\eqref{EqGammaRMTDevelopped} leads to $(4!)^2=576$ different terms. This average can be expressed as~\cite{mello90,brouwer96}
\begin{multline}
\overline{(U_{a_1b_1} \dots U_{a_4b_4})(U_{\alpha_1\beta_1}^* \dots U^*_{\alpha_4\beta_4})}\\= \sum_{P,P'}V_{PP'}\prod_{j=1}^4\delta_{a_j\alpha_{P(j)}}\delta_{b_j\beta_{P'(j)}},
\label{EqUnitary}
\end{multline}
where $P,P'$ are permutations of $\{1,2,3,4\}$, and the weights $V_{PP'}$ depend on the cycle structure of $P^{-1}P'$ only: $V_{PP'}=V_{c_1 \dots c_k}$, with $c_j$ the lengths of disjoint cyclic permutations in $P^{-1}P'$ ($\sum_{j=1}^k c_j=4$). In the limit $N\gg1$, it can be shown that~\cite{brouwer96}
\be
V_{c_1 \dots c_k}=\prod_{j=1}^kV_{c_j}, \;\;\; \text{with} \;\;\; V_c=\frac{(-1)^{c-1}(2c-2)!}{c(c-1)!^2} \frac{1}{N^{2c-1}}.
\ee
Hence, although we have $576$ terms in Eq.~\eqref{EqUnitary}, we only have five different weights in the limit $N\gg1$:
\begin{align}
V_{1111}=V_1^4=\frac{1}{N^4},\; V_{112}=V_1^2V_2=-\frac{1}{N^5},\nonumber \\
V_{13}=V_1V_3=\frac{2}{N^6},\; V_{22}=V_2V_2=\frac{1}{N^6},\; V_4&= -\frac{5}{N^7}.
\end{align}
Let us first consider the terms in Eq.~\eqref{EqGammaRMTDevelopped} associated to the weight $V_{1111}=1/N^4$. They are obtained for $P=P'$ and correspond to the $4!=24$ terms found by treating $U$ as a Gaussian random matrix. A simple calculation gives
\begin{widetext}
\be
\Gamma_{ba}^{(\text{Gaussian} \; U)} = (1+\delta_{ab})\frac{\overline{\text{Tr}(R)^2}}{N^3}+(2+4\delta_{ab})\frac{\overline{\text{Tr}(R)^2}}{N^4}+(1+\delta_{ab})\frac{\overline{\text{Tr}(R^2)}}{N^3} +\mathcal{O}\left(\frac{1}{N^3}\right).
\label{EqGammaGaussianU} 
\ee
\end{widetext}
 In this expression, we put aside the term $(5+9\delta_{ab}) \overline{\text{Tr}(R)^2}/N^4= \mathcal{O}(1/N^3)$, which can be neglected for $N\gg 1$ since $\overline{\text{Tr}(R^2)}=\mathcal{O}(N) $ and $\overline{\text{Tr}(R)^2}=\mathcal{O}(N^2)$.  It is instructive to compare this result with the one obtained by treating $r$ as a Gaussian random matrix, as was done in the previous section:
\be
\Gamma_{ba}^{(\text{Gaussian} \; r)} = (1+\delta_{ab})\frac{\overline{\text{Tr}(R)^2}}{N^3}+(2+4\delta_{ab})\frac{\overline{\text{Tr}(R)^2}}{N^4}.
\label{EqGammaGaussianR} 
\ee
The difference between Eq.~\eqref{EqGammaGaussianR} and Eq.~\eqref{EqSolGammaC1} comes from the waveguide geometry where $W<\ell$. Since the mode spacing in the waveguide is larger than the width $\sim1/\ell$ of the function $F(q)$, only the central component of $F(q)$ is probed. The result~\eqref{EqGammaGaussianR} can be recovered from Eq.~\eqref{EqSolGammaC1} by using $F(q)\simeq F(0)\delta_{q,0}$, with $F(0)=\overline{\text{Tr}(R)}/N^2$. In addition, the comparison of Eq.~\eqref{EqGammaGaussianU} and Eq.~\eqref{EqGammaGaussianR} reveals that the Gaussian approximation for $r$ does not capture the third term of weight $\overline{\text{Tr}(R^2)}/N^3 \sim 1/N^2$, which is of the same order as the second one, $\overline{\text{Tr}(R)^2}/N^4 \sim 1/N^2$. This indirectly shows that the result~\eqref{EqSolGammaC1}, based on a Gaussian approximation for $r$, misses scattering contributions which are of the same order as its subleading components.

Let us now consider the remaining $576-24=552$ terms in Eq.~\eqref{EqGammaRMTDevelopped}. According to Eq.~\eqref{EqUnitary}, each of the $24$ Gaussian terms can give rise to $23$ new terms by considering $P'\neq P$. Among all of them, we are interested in the leading ones only. They are those which yield contributions of the same order as the last two terms of Eq.~\eqref{EqGammaGaussianU}. Such contributions must fulfill three conditions.  First, they must have a weight $V_{112}=-1/N^5$, which corresponds to $P'$ different from $P$ by a permutation of two indices only: this leaves $6$ possibilities instead of $23$ for each Gaussian diagram. Second, they must preserve one free summation $\sum_{a'} =N$ in Eq.~\eqref{EqGammaRMTDevelopped}. Only $4$ Gaussian terms satisfy this property. We are thus left with $4\times 6 = 24$ new terms. Finally, we select contributions which have a weight $\overline{Tr(R)^2} \sim N^2$ and disregard those of weight $\overline{Tr(R^2)} \sim N$. In this way, we end up with $8$ new terms only, 
\be
\Gamma_{ba}^{(\text{correction})} =-4(1+\delta_{ab})\frac{\overline{\text{Tr}(R)^2}}{N^4},
\label{EqGammaCorr} 
\ee
which turn out to compensate most of the sub-leading Gaussian contributions [see Eq.~\eqref{EqGammaGaussianR}]. By combining Eq.~\eqref{EqGammaGaussianU} with  Eq.~\eqref{EqGammaCorr}, we finally conclude that
\begin{multline}
\Gamma_{ba} =(1+\delta_{ab})
\frac{\overline{\text{Tr}(R)^2}}{N^3}
+(1+\delta_{ab})
\frac{\overline{\text{Tr}(R^2)}}{N^3}\\
-2\frac{\overline{\text{Tr}(R)^2}}{N^4}
 +\mathcal{O}\left(\frac{1}{N^3}\right).
\label{EqGammaFull} 
\end{multline}
Interestingly, this result predicts a 2p-CBS enhancement slightly larger than $2$:
\be
\frac{\Gamma_{a a}}{\Gamma_{b \neq a}}=2\left(1+\frac{1}{N}\right)  +\mathcal{O}\left(\frac{1}{N^2}\right).
\ee
In order to test the validity of the prediction~\eqref{EqGammaFull},  we computed numerically Eq.~\eqref{EqDefGammaRMT} with the   decomposition~\eqref{EqRSVD}. For clarity, we considered the limit of large optical thickness $L\gg \ell$, where the diagonal matrix $R$ can be replaced by the identity. In that situation, Eq.~\eqref{EqGammaFull} gives $\Gamma_{b\neq a}=(1-1/N)/N$ and $\Gamma_{a a}=2/N$. Figure~\ref{FigRMT}(a) shows that these predictions are in excellent agreement with numerical results obtained with $10^5$ realizations of the random unitary matrix $U$. We also compared these results for 2p-CBS  with the RMT prediction for $R_{ba}=\overline{\vert r_{ba} \vert^2}$ characterizing 1p-CBS. The latter is much easier to compute  than $\Gamma_{ba}$ because it involves $(2!)^2=4$ terms instead of $(4!)^2=576$. The average in the unitary group gives 
\be
R_{ba}=(1+\delta_{ab})\left(1-\frac{1}{N}\right)
\frac{\overline{\text{Tr}(R)}}{N^2},
\ee
and thus  $R_{aa}/R_{b\neq a}=2$ for any $N$. This prediction also agrees with numerical simulations, as  shown in Fig.~\ref{FigRMT}(b).  
\begin{figure}[t]
\centering
\includegraphics[width=1\linewidth]{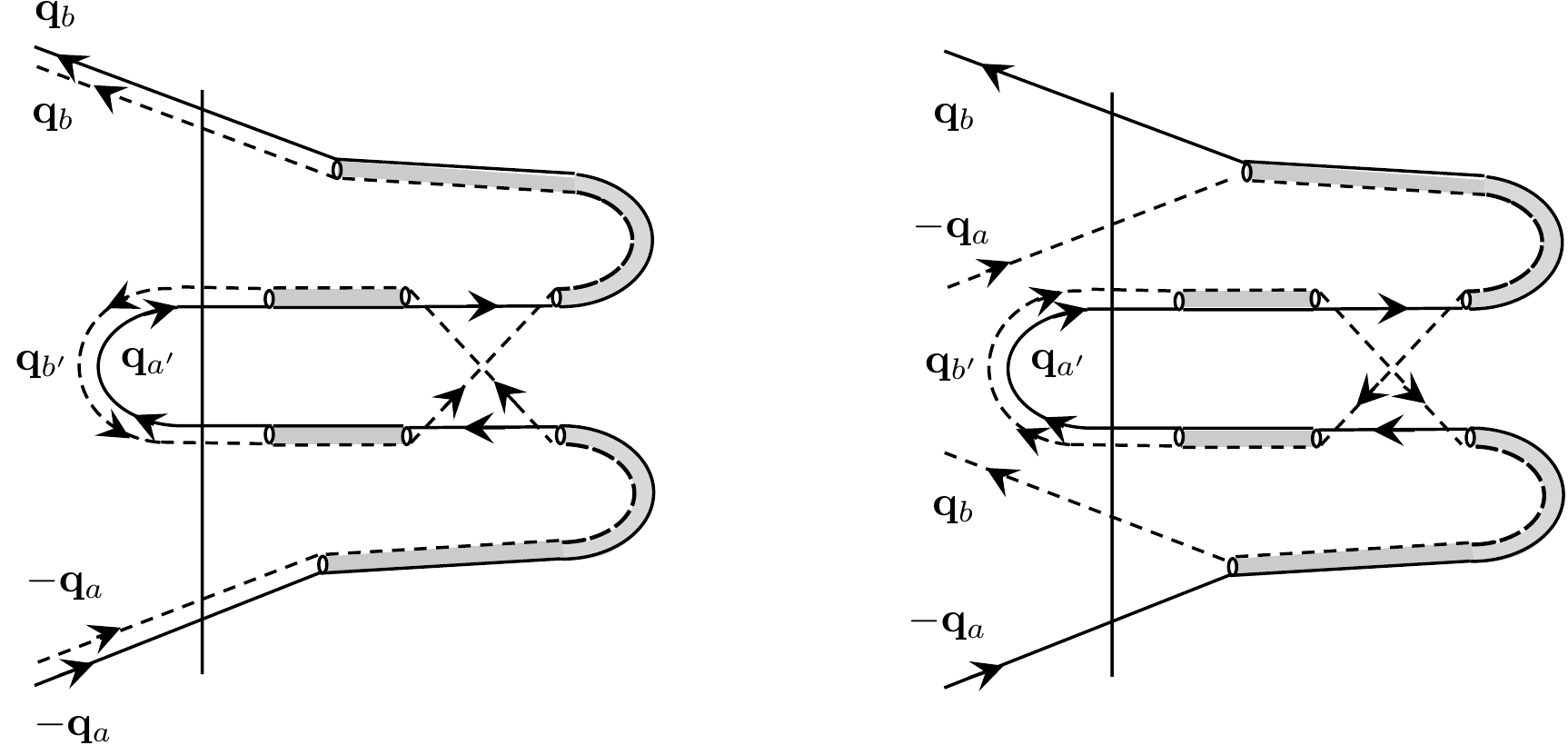}
\caption{Microscopic representation of non-Gaussian diagrams contributing to $\Gamma_{ba}$. In the RMT framework, the weights of these diagrams are $\overline{\text{Tr}(R^2)}/N^3$ and $\overline{\text{Tr}(R^2)} \delta_{ab}/N^3$, respectively. Each of them involves an exchange of diffusive partners in the bulk of the medium, mathematically described by a Hikami box (the latter contains three possible ways to exchange partner, instead of one represented here).}
\label{FigDiag3}
\end{figure} 

The comparison of the rigorous expansion~\eqref{EqGammaFull} with the Gaussian result~\eqref{EqGammaGaussianR} reveals that the first non-Gaussian contribution corresponds to the terms $(1+\delta_{ab})\overline{\text{Tr}(R^2)}/N^3$,
 generated by expressing the average in Eq.~\eqref{EqGammaRMTDevelopped}  as a product of pairs of the form $\overline{U_{\alpha \beta}U^*_{\gamma \delta}}$. Keeping track of these pairings allows to
 find their microscopic diagrammatic representation, in terms of diffusons, cooperons, and Hikami boxes. We obtain the two classes of diagrams shown in Fig.~\ref{FigDiag3}, which can be interpreted as the multiple scattering counterparts of the non-Gaussian terms already identified in the double-passage configuration (see middle column of Fig.~\ref{FigDiagDouble}). They both involve exchanges of field partners before and after the virtual back-reflection associated to the maximally entangled pair injection. They are similar to the diagrams standing for weak-localization corrections to diffusive transport, and are thus expected to have a weight $1/k\ell$ with respect to the dominant Gaussian terms~\eqref{EqSolGammaC1Approx}.  This is perfectly consistent with the RMT approach in which they have a relative weight $1/N$ [see Eq.~\eqref{EqGammaFull}]. Hence, the RMT approach allows us to conclude that the result~\eqref{EqSolGammaC1Approx} is rigorous as long as $k\ell \gg1$. All subleading contributions, such as the ones represented in Fig.~\ref{FigDiag3}, are expected to become important close to the onset of Anderson localization.

\section{E\lowercase{xtended simulation results}}

\begin{figure}[htb!]
 \centering
 \includegraphics[width=0.7\columnwidth]{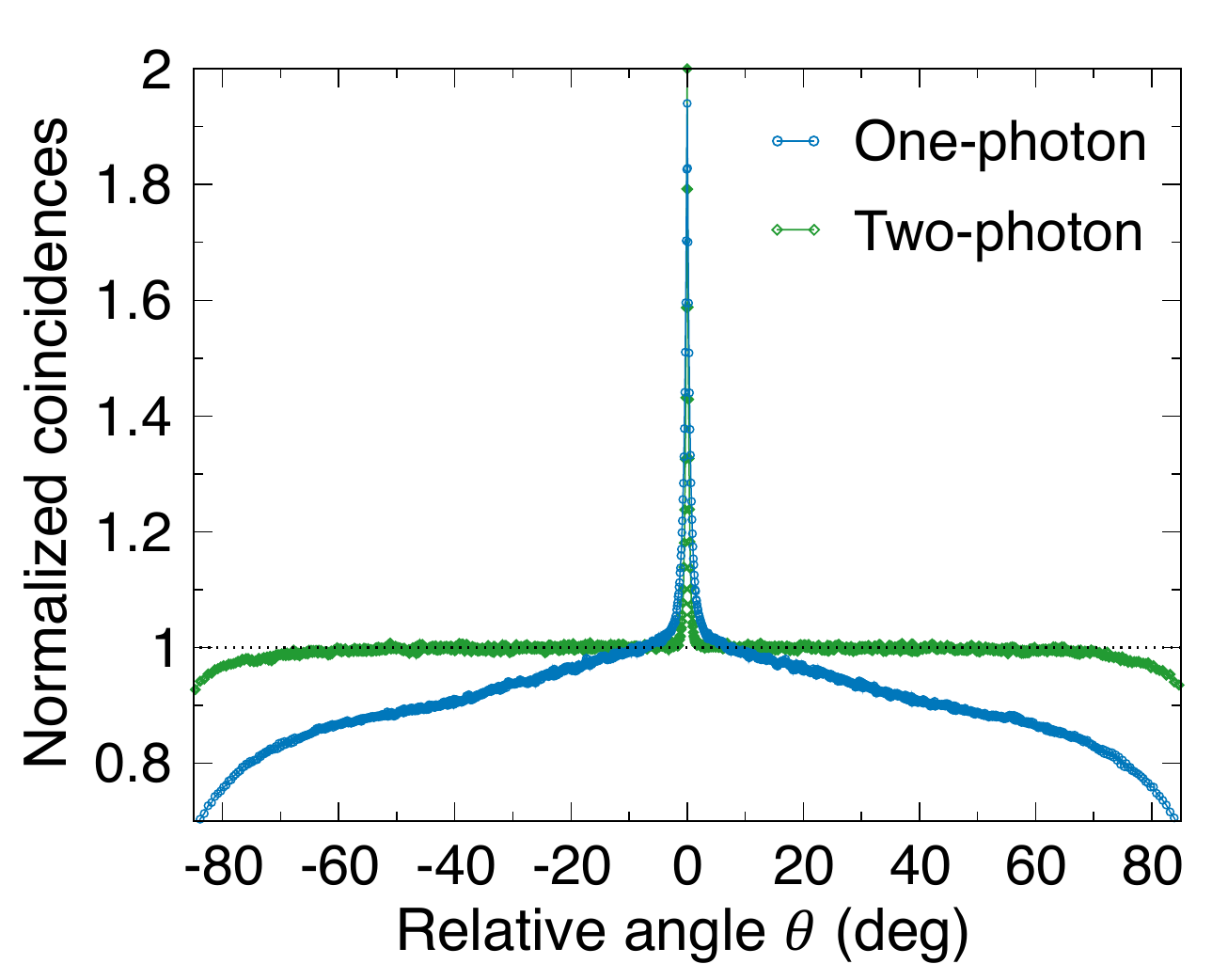}
 \caption{One-photon and two-photon coincidence rates in reflection plotted over the full angular range. Same data as Fig.~4(b) in main text except with larger angular range.}
 \label{fig:sim_full_range}
\end{figure}

Figure~\ref{fig:sim_full_range} shows the one-photon and two-photon coincidence rates from numerical simulations of disordered media, plotted over the full angular range. These curves are used to determine the normalization factors $R_0$ and $\Gamma_0$.

\begin{figure*}[hbt!]
 \centering
 \includegraphics[width=1.5\columnwidth]{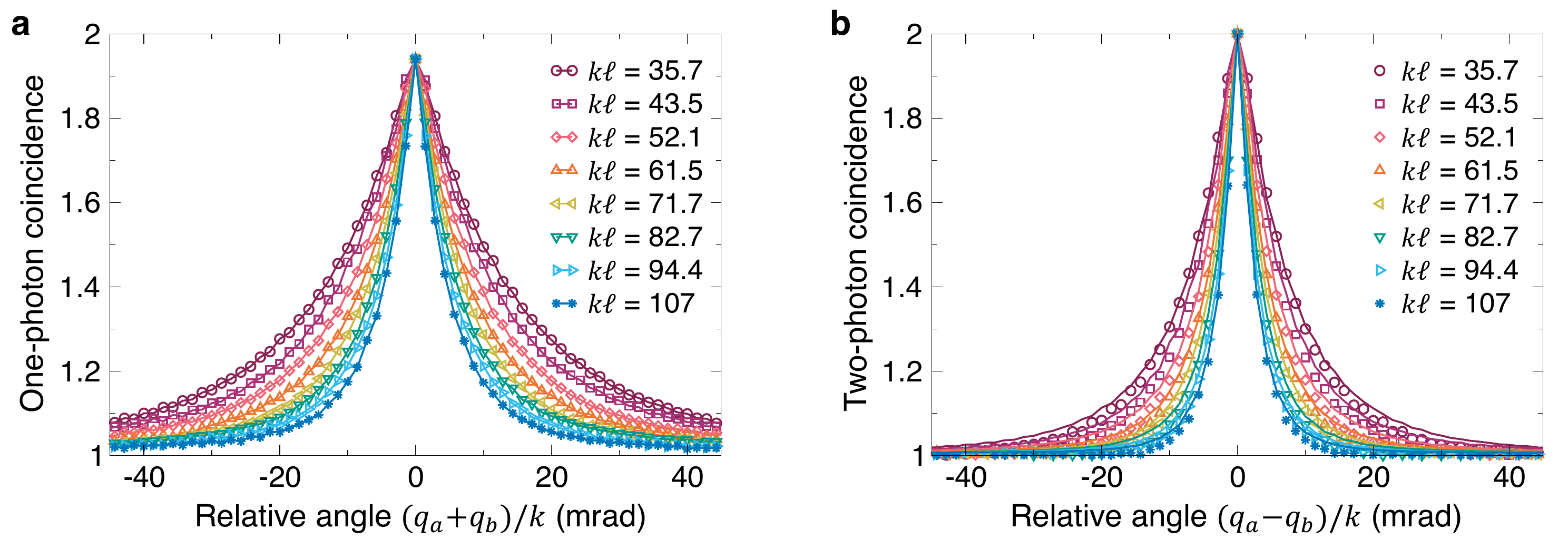}
 \caption{One-photon and two-photon CBS for different transport mean free paths. \textbf{a}, Normalized one-photon coincidence rates, $R_{ba}/R_0$. \textbf{b}, Normalized two-photon coincidence rates, $\Gamma_{ba}/\Gamma_0$ in symbols and $1 + [(R_{ba}/R_0)-0.94]^2$ in solid lines. The transport mean free path $\ell$ is varied by changing the number density $\rho$ of dielectric cylinders.
}
 \label{fig:all_lt}
\end{figure*}

Figure~\ref{fig:all_lt} shows the numerical CBS data for different transport mean free paths. These curves are used to determine the FWHM reported in Fig.~4(c) of the main text.

\section{F\lowercase{isher information in 1p-\uppercase{CBS} and 2p-\uppercase{CBS}}}

The objective of this section is to evaluate the Fisher information in 1p-CBS and 2p-CBS experiments, in the presence of Poisson and speckle noises. For a set of measured data $\vec{X}$ which depend on some unknown parameters $\bm{\theta}=\{\theta_1, \theta_2, \dots \}$, the Fisher information matrix $\mathcal{F}$ sets a lower bound to the variance of any estimator $\hat{\theta}_i$ of $\theta_i$, in the form~\cite{kay93}
\be
\text{Var}\left( \hat{\theta}_i \right) \ge [\mathcal{F}^{-1}]_{ii}.
\ee
The term $[\mathcal{F}^{-1}]_{ii}$ is called the Cramer--Rao lower bound (CRLB) on the parameter $\theta_i$. It is expressed in terms of the inverse of the Fisher matrix $\mathcal{F}$, whose size is equal to the number of unknown parameters to be estimated. The elements of  $\mathcal{F}$ are
\be
\mathcal{F}_{ij}=\int d\vec{X}\frac{\partial_{\theta_i}p(\vec{X} \vert \bm{\theta} )\partial_{\theta_j}p(\vec{X} \vert \bm{\theta} )}{p(\vec{X} \vert \bm{\theta} )},
\label{EqF}
\ee
where the conditional probability $p(\vec{X} \vert \bm{\theta} )$ is the probability to measure $\vec{X}$. In an optical experiment, the expression of  $p(\vec{X} \vert \bm{\theta} )$ depends on the illumination, the scattering system, and the detection scheme.

 Let us first discuss the 1p-CBS experiment. We represent  $\vec{X}$ as the set of numbers of photon $n_k$ recorded for each position $b$ of the detector $D_b$, $\vec{X}=\{n_1, \dots, n_M \}$ where $M$ is the total number of scanned positions, and we assume the successive positions to be separated by a distance larger than a speckle grain size. In that case, the numbers $n_b$ are independent of each other, so that  $p(\vec{X} \vert \bm{\theta} )=\prod_{b=1}^Mp(n_b \vert \bm{\theta} )$, and Eq.~\eqref{EqF} becomes
 \be
 \mathcal{F}_{ij}=\sum_{b=1}^M \mathcal{F}^{(b)}_{ij},
 \label{EqFSum}
 \ee
with $\mathcal{F}^{(b)}_{ij}$ the Fisher information for position $b$,
\be
\mathcal{F}^{(b)}_{ij}=\int dn_b\frac{\partial_{\theta_i}p(n_b \vert \bm{\theta} )\,\partial_{\theta_j}p(n_b\vert \bm{\theta} )}{p(n_b \vert \bm{\theta} )}.
\label{EqF4}
\ee 
If we were considering photon-noise only, we would use the Poisson distribution, 
$p(n_b\vert \bm{\theta} )=\bar{n}_b^{n_b}e^{-\bar{n}_b}/n_b!$, with $\bar{n}_b$ the mean photon number proportional to the CBS intensity profile. However, in the presence of speckle noise, $\bar{n}_b$ strongly fluctuates from one disorder realization to another. Hence, the probability to collect $n_b$ photons at position $b$ takes the form~\cite{goodman65}
 \be
p(n_b\vert \bm{\theta} )=\int d\bar{n}_b \frac{\bar{n}_b^{n_b}}{n_b!}e^{-\bar{n}_b} p(\bar{n}_b\vert \bm{\theta} ),
\label{EqPjoint}
\ee
where $p(\bar{n}_b\vert \bm{\theta} )$ depends on the number of disorder realizations. For a single disorder configuration, the latter is given by the Rayleigh distribution, $p(\bar{n}_b\vert \bm{\theta} )=e^{-\bar{n}_b/\left<\bar{n}_b \right>}/\left<\bar{n}_b \right>$, whereas for $N_r\gg1 $ configurations, we obtain a Gamma distribution of order $N_r$. This can be shown by considering the random variable $\bar{n}_b=\sum_{i=1}^{N_r}\bar{n}^{(i)}_b/N_r$, where each $\bar{n}^{(i)}_b$ obeys the Rayleigh distribution. Using the fact that the characteristic function of $\sum_{i=1}^{N_r}\bar{n}^{(i)}_b$ is the product of the characteristic functions of each $\bar{n}^{(i)}_b$, we get~\cite{goodman2015statistical}
\be
p(\bar{n}_b\vert \bm{\theta} )=\frac{N_r^{N_r}}{\Gamma(N_r)} \frac{\bar{n}_b^{N_r-1}e^{-N_r\bar{n}_kb/\left<\bar{n}_b \right>}}{\left<\bar{n}_b \right>^{N_r}},
\label{EqGamma}
\ee
which converges to a Gaussian of root mean square $\sigma=\left<\bar{n}_b \right>/\sqrt{N_r}$ for $N_r\gg 1$. With Eq.~\eqref{EqGamma}, the distribution~\eqref{EqPjoint} becomes
\be
p(n_b\vert \bm{\theta} )= \frac{\Gamma(N_r+n_b)}{\Gamma(N_r)\Gamma(n_b+1)}\frac{N_r^{N_r}\left<n_b \right>^{n_b}}{(N_r+\left<n_b \right>)^{N_r+n_b}},
\label{EqPoissonSpeckle}
\ee
where $\left<n_b \right>\equiv\left<\bar{n}_b \right>$. We can now evaluate the Fisher information at position $b$. Noting that
\be
\partial_{\theta_i}p(n_b\vert \bm{\theta} )=\frac{N_r(n_b-\left<n_b \right>)}{\left<n_b \right>(N_r+\left<n_b \right>)}p(n_b\vert \bm{\theta} )\,\partial_{\theta_i}\langle n_b \rangle,
\ee
Eq.~\eqref{EqF4} becomes
\be
\mathcal{F}^{(b)}_{ij}=
\left[\frac{N_r}{\left<n_b \right>(N_r+\left<n_b \right>)}\right]^2 \text{Var}(n_b) \partial_{\theta_i}\langle n_b \rangle  \partial_{\theta_j}\langle n_b \rangle,
\label{FisherPoissonSpeckle}
\ee
where $\text{Var}(n_b)=\left< n_b^2\right>- \langle n_b \rangle^2$ is the variance of the photon number. According to the decomposition~\eqref{EqPjoint}, the latter is simply the sum of Poisson and speckle contributions, $\text{Var}(n_b)=\left< n_b\right> + \left< n_b\right>^2/Nr$. Hence, we finally express the Fisher information as
\be
\mathcal{F}^{(b)}_{ij}= \frac{N_r}{N_r+\left< n_b\right>} \frac{\partial_{\theta_i}\langle n_b \rangle  \partial_{\theta_j}\langle n_b \rangle}{\left< n_b\right>}.
\label{FisherPoissonSpeckle2}
\ee

Let us  apply previous formula to estimate the transport mean free path $\ell$, from measurements of the 1p-CBS performed in the limit of large photon number where speckle noise dominates ($\left< n_b\right> \gg N_r$). The number of detected photon $\left< n_b\right>$ being proportional to $R_{ba}$, we find 
\be
\mathcal{F}^{\text(1p)}_{\ell \ell}=N_r \sum_{\vec{q}_b} \left[ \frac{\partial_\ell R_{ba}}{R_{ba}} \right]^2,
\label{FisherOnePhoton}
\ee   
where $R_{ba}=F(0) + F(\vert \vec{q}_a+\vec{q}_b \vert)$ and $F(q)$ is given by Eq.~\eqref{EqFctCBS}. The sum~\eqref{FisherOnePhoton} could be evaluated explicitly to get an expression of the CRLB in terms of $\ell$. More interestingly, we will establish below a relation between  $\mathcal{F}^{\text(1p)}_{\ell \ell}$ and $\mathcal{F}^{\text(2p)}_{\ell \ell}$, which holds independently of the precise expression of $F(q)$. 

In a two-photon speckle experiment, we measure the coincidence rate $\Gamma_{ba}$ of two detectors counting photons in directions $\vec{q}_a$ and $\vec{q}_b$. The number of coincidences measured during the acquisition time, $N_{ba} \propto \Gamma_{ba}$,  does not obey the Rayleigh statistics in general. However, in the special case of a maximally entangled input state made of $N\gg 1$ modes, it can be shown that $p(N_{ba})$ is exponentially distributed. This means that we can apply the result~\eqref{FisherPoissonSpeckle2}  to our 2p-CBS experiment, with $N_{ba}$ replacing $\left< n_b \right>$. In particular, for $N_{ba} \gg N_r$, we get
\be
\mathcal{F}^{\text(2p)}_{\ell \ell}=N_r \sum_{\vec{q}_b} \left[ \frac{\partial_\ell \Gamma_{ba}}{\Gamma_{ba}} \right]^2,
\label{FisherOnePhoton}
\ee 
where  $\Gamma_{ba}=F(0)^2 + F(\vert \vec{q}_a-\vec{q}_b \vert)^2$ (see Sec~\ref{SecTheory} for details). For a given detection direction $\theta$, the Fisher information ratio $\mathcal{F}^{\text(2p)}_{\ell \ell}/\mathcal{F}^{\text(1p)}_{\ell \ell}$ takes the form
\be
\frac{\mathcal{F}^{\text(2p)}_{\ell \ell}}{\mathcal{F}^{\text(1p)}_{\ell \ell}}=4 \frac{1+F(0)/F(2k\text{sin}\theta/2)}{1+\left[F(0)/F(2k\text{sin}\theta/2)\right]^2}.
\ee
It it maximal at the center of the CBS peak ($\theta=0$), where it takes the value $\mathcal{F}^{\text(2p)}_{\ell \ell}/\mathcal{F}^{\text(1p)}_{\ell \ell}=4$, irrespectively of $F(q)$. This implies that the CRLB on $\ell$ is reduced by a factor $4$ when using 2p-CBS produced by the EPR state $\ket{\psi} = \sum_{i=1}^{N} \hat{c}_{\mathbf{q}_i}^{\dagger}\hat{c}_{-\mathbf{q}_i}^{\dagger}\ket{0}/\sqrt{2N}$.


\bibliography{SI_refs}